\documentclass[letterpaper]{article} 
\usepackage{aaai2026}  
\usepackage{times}  
\usepackage{booktabs}
\usepackage{longtable}
\usepackage{caption}
\usepackage{rotating}
\usepackage{pdflscape}
\usepackage{array}
\usepackage{helvet}  
\usepackage{courier}  
\usepackage[hyphens]{url}  
\usepackage{graphicx} 
\usepackage{natbib}  
\usepackage{caption} 
\usepackage{algorithm}
\usepackage{algorithmic}

\usepackage{xcolor}
\newcommand{\answerYes}[1]{\textcolor{blue}{#1}} 
\newcommand{\answerNo}[1]{\textcolor{teal}{#1}} 
\newcommand{\answerNA}[1]{\textcolor{gray}{#1}}

\usepackage{newfloat}
\usepackage{listings}
\DeclareCaptionStyle{ruled}{labelfont=normalfont,labelsep=colon,strut=off} 
\floatstyle{ruled}
\newfloat{listing}{tb}{lst}{}
\floatname{listing}{Listing}
\title{Detecting and Enhancing Intellectual Humility in Online Political Discourse}

\author {
    Samantha D'Alonzo\textsuperscript{\rm 1},
    Rachel Chen\textsuperscript{\rm 1*},
    Weidong Zhang\textsuperscript{\rm 1*},
    Melody Yu\textsuperscript{\rm 1},
    Jasmine Mangat\textsuperscript{\rm 1},
    Ivory Yang\textsuperscript{\rm 2},
    Weicheng Ma\textsuperscript{\rm 3},
    Martin Saveski\textsuperscript{\rm 4},
    Soroush Vosoughi\textsuperscript{\rm 2},
    Nabeel Gillani\textsuperscript{\rm 1, 5}
}
\affiliations {
    \textsuperscript{\rm 1}College of Arts, Media and Design, Northeastern University\\
    \textsuperscript{\rm 2}Department of Computer Science, Dartmouth College\\
    \textsuperscript{\rm 3}School of Engineering and Computer Science, Oakland University\\
    \textsuperscript{\rm 4} Information School, University of Washington\\
    \textsuperscript{\rm 5}D'Amore-McKim School of Business, Northeastern University\\
    \{dalonzo.sa, chen.rachel, zhang.weid, yu.melo, j.mangat, n.gillani\}@northeastern.edu\\
    \{ivory.yang.gr, soroush.vosoughi\}@dartmouth.edu\\
    weichengma@oakland.edu,
    msaveski@uw.edu
}

\begin{document}
\maketitle
\renewcommand{\thefootnote}{*}
\footnotetext{Contributions of equal importance}
\renewcommand{\thefootnote}{\arabic{footnote}}
\begin{abstract}
Intellectual humility (IH)---a recognition of one's own intellectual limitations---can reduce polarization and foster more understanding across lines of difference. Yet little work explores how IH can be systematically defined, measured, evaluated, and enhanced in spaces that often lack it the most: online political discussions. In this paper, we seek to bridge these gaps by exploring two questions: 1) how might pre-existing levels of IH influence future expressions of IH during online political discourse? and 2) can online interventions enhance IH across different political topics and conversational environments? To pursue these questions, we define a codebook characterizing different dimensions of IH and intellectual arrogance (IA) and have researchers use it to annotate several hundred Reddit posts, which we then use to develop and validate a classifier to support IH analysis at scale. These tools subsequently enable two key contributions: i) an observational data analysis of how IH varies across different political discussions on Reddit, which reveals that more/less IH environments tend to contain future posts of a similar nature, and ii) a randomized control trial evaluating strategies for nudging discussion participants to demonstrate more IH in their posts, which reveals the possibility of enhancing IH in online discussions across a range of contentious topics. Our findings highlight the possibility of measuring and increasing IH online without necessarily reducing engagement.
\end{abstract}

\section{Introduction}
For decades, scholars across disciplines have studied intellectual humility (IH): the practice of acknowledging one's own intellectual limitations \cite{whitcomb_intellectual_2017}. Recent work has found that greater IH is often associated with reductions in political hostility~\cite{smith_you_2023}, less political polarization and ``myside'' bias~\cite{stroud_intellectual_2025, bowes_stepping_2022}, and enhanced tolerance between religious groups \cite{hook_intellectual_2017}. Promoting IH has also been shown to improve student learning \cite{porter_teachers_2024} and lead to more open communication in certain corporate settings \cite{niu_influence_2025}. These promising associations highlight IH's foundational role in fostering peace, cooperation, and other constructive practices across lines of difference and have contributed to increased research on how to promote IH, including in online settings~\cite{welker_online_2023}, like search and social media platforms~\cite{rieger_potential_2024,van_loon_designing_2024}. 

Yet efforts to foster greater IH on social media platforms face strong headwinds. The emphasis on engagement optimization for social media platforms has contributed to the amplification of antagonism and affective polarization \cite{tornberg_modeling_2021}. Many believe that unbridled engagement optimization has also contributed to increases in the spread of misinformation \cite{m_spreading_2016}. Under these conditions, social media environments can create negative feedback loops, where antisocial (e.g., toxic) conversations and environments breed more toxicity~\cite{saveski2021,xia2020}---demonstrating how such antisocial behaviors are driven not only by personal disposition, but also, characteristics of conversation environments \cite{cheng2017}. 

Promoting IH on social media falls under broader efforts to redesign social media interfaces and algorithms in ways that promote prosocial virtues like civility, empathy, and cooperation. Several of these efforts have received increased attention in recent years~\cite{dorr2025, matias2019, jia_embedding_2024}, with many focused on enhancing civility and reducing toxicity. While these are important virtues to promote, both civility and lack of toxicity can be seen as minimum requirements for discourse. IH, on the other hand, is a higher discourse standard to strive for. 

Civility may involve tempering the expression of appropriate attitudes, such as anger and indignation. As a result, citizens might conform to unjust social arrangements rather than improve them \cite{ward_democratic_2017}. Civility may also favor avoiding certain, difficult topics, such as racism, instead of actively discussing and addressing them \cite{davis_role_2021}. IH, on the other hand, involves active engagement with different opinions and one's own limitations in a way that could lead to productive dialogue on important democratic topics rather than avoidance or surface-level, polite acknowledgment of differences. Similarly, toxicity, as defined by PerspectiveAPI \cite{jigsaw_perspective_nodate}, involves ``rude, disrespectful, or unreasonable comment[s] that [are] likely to make people leave a discussion''. Eliminating such language may mean people are more likely to stay engaged in discussion, but it does not increase the likelihood that discussions about divisive topics yield beneficial outcomes. 

IH is also distinct from both tone and sentiment, which are attributes of speech, rather than normative values expressed through speech. A crucial aspect of IH that distinguishes it from civility is its capacity to express negative sentiment or a serious tone in a manner conducive to continued conversation. Therefore, IH is orthogonal to tone and sentiment.

Relative to civility and toxicity, there is also a lack of research on IH interventions, despite the foundational nature and distinct importance of this particular virtue. Additionally, several studies have uncovered the effects that environmental toxicity and other antisocial attributes can have on the posting behavior and content of individuals in those environments~\cite{saveski2021,xia2020}, but comparatively less work has explored the extent to which thoughtfully-designed interventions might overcome these environmental barriers to break the negative feedback loops they otherwise threaten to perpetuate. In this study, we seek to bridge these gaps by investigating how environments influence levels of IH, and how increasing IH through exogenous interventions might, in turn, impact discussion environments in social media settings. In particular, we ask the following overarching research questions:

\begin{itemize}
    \item RQ1: Do pre-existing levels of intellectual humility in online environments influence the expression of intellectual humility in future conversations?
    \item RQ2: Can online interventions enhance the prevalence of intellectually humble discourse across different discussion topics and conversational environments?
\end{itemize}

To pursue these questions, we build on~\citet{guo_computational_2024} to operationalize a definition of IH/IA in the context of online political discourse through the development of a codebook. We use this codebook to annotate a dataset of 359 Reddit posts from political subreddits. We then use this codebook and set of human annotations to construct an LLM-based IH classifier in order to enhance our ability to detect and measure IH at scale, achieving promising performance on a task that is difficult even for humans. This tooling enables our two key contributions. First, we use the classifier to explore RQ1 by analyzing existing Reddit data from political subreddits from March through May 2024 to identify associations between levels of IH in subreddits and levels of IH demonstrated by those subsequently participating in them. And second, we further explore RQ1 and RQ2 through a lab-based randomized experiment to evaluate the extent to which a novel web-based intervention can increase levels of intellectual humility in online discussions. The lab-based experiment extends the findings from the secondary Reddit data analysis to identify causal effects of environment on future discussions. It also demonstrates the intervention's potential to enhance IH in online discourse, even in environments that demonstrate more IA. Interestingly, the intervention also does not reduce explicit engagement (measured by the number of times participants post), suggesting that fostering greater IH in online public discourse need not necessarily come at the expense of social media companies' ``North Star'' metrics. 

While prior studies have developed IH classifiers~\cite{guo_computational_2024}, analyzed how the attributes of a discussion environment can impact future discussions~\cite{saveski2021,xia2020,stroud_intellectual_2025}, and investigated how just-in-time interventions in discussion spaces might reduce harmful / promote more virtuous content like~\cite{katsaros2021,Argyle2023_democratic_discourse}, our study offers a novel contribution by 1) focusing on the important virtue of IH, 2) exploring how analyses and impacts might vary across discussion topics and types of environments, and 3) demonstrating how these different activities---conceptualization of IH, classification, analysis, and intervention---can be integrated to foster greater IH online. To support researchers interested in building upon our work, we release our annotated datasets, classifier, and code for the online intervention environment\footnote{\url{https://github.com/Plural-Connections/ih-political-discourse}}. Together, our contributions highlight both challenges and opportunities for enhancing IH in online political discourse.
 
\section{Related Work}

\subsection{Measuring Toxicity and Contagion Effects} Many scholars have studied how online environments influence users, and mainly focusing on, but not limited to, toxicity.~\citet{saveski2021} offered a structural perspective, showing that features like reply depth and social network sparsity predict whether a conversation will escalate into toxicity. This work builds on prior studies like ~\cite{liu2018instagramhostility,hessel-lee-2019-somethings} showing that existing conversation dynamics can be used to forecast future discourse features. Other studies like ~\citet{xia2020} have found that both a comment's own toxicity and the ability to elicit toxic replies is shaped by the conversational context. Others have shown that trolling is not just a matter of personal disposition; instead, exposure to prior toxic comments can increase online discussants' propensity to perpetuate toxic discourse \cite{cheng2017}. On the other hand, pro-social environments are not a panacea. Research has found that social media users who identify with exclusionary themes or practices (like the “Manosphere” on Reddit) retain their distinctive linguistic styles even in unrelated Subreddits---suggesting limited responsiveness to context \cite{aggarwal2024}. Still, well-articulated and enforced norms can reduce the spread of harmful content~\cite{park_how_2023}.

\subsection{Interventions to Improve Online Discourse} Recognizing the impact that discussion environments can have on future discussion quality, several studies have sought to intervene to prevent potentially toxic or harmful content from being posted. Recent quasi-experimental work found that positive feedback on Reddit can significantly enhance the quality of future posts~\cite{Lambert2024PositiveReinforcement}. Prior experimental work has shown that simply reminding people of discussion norms can reduce antisocial behavior like harassment~\cite{matias2019}. Additionally, intervening during the time of content creation---both through simple nudges prompting reconsideration of possibly harmful posts~\cite{katsaros2021} and dynamic suggestions for updating posts produced by large language models (LLMs)~\cite{Argyle2023_democratic_discourse}---can also improve discussion quality and other downstream outcomes. Fewer (though a growing number of) interventions have focused specifically on enhancing IH in online discourse.  Recently,~\cite{van_loon_designing_2024} defined IH using  PerspectiveAPI~\cite{jigsaw_perspective_nodate} and discovered that social norms (operationalized through badges) extolling users for demonstrating ``open-mindedness'' might enhance IH online. Still, not all uses of AI in online discussion spaces positively impact discourse, as~\cite{MollerRomeroJurgensAiello2025} shows.

Our paper builds on many of these individual studies to offer one integrated approach to defining, measuring, evaluating, and enhancing IH in online political discourse, further illustrating how these enhancements might remain robust to different discussion topics and environments.

\section{Codebook and Classifier Development}
We adapted the codebook developed by \citet{guo_computational_2024}, which identified Intellectual Humility (IH) and Intellectual Arrogance (IA) in conversations about religion, to identify both IH and IA in political discussions on Reddit. Each broad label, IH or IA, had a number of sub-labels associated with it. For each sub-label and definition, we chose to either keep the sub-label as is, remove it, or adapt it for our use case. We also generated two new politics-specific IA sub-labels: `Polarizing or Tribalistic Language' and `Overinflated Expertise'. These decisions were informed by literature reviews on IH in political discourse and an assessment of differences between political discourse and religious discourse. Specific justifications for each decision to keep, cut, shift, or generate a new label can be found in Appendix A.

Three expert human annotators (two undergraduates and one graduate student, described in Appendix A.1) then iteratively applied the codebook to annotate a random sample of Reddit posts. We sampled 350 posts from 8 popular subreddits across the Left-Right ideological spectrum focused on US Politics (r/Conservative, r/Libertarian, r/Liberal, r/PoliticalDiscussion, r/politics, r/conservatives, r/Republican, r/uspolitics)  over a three-month time horizon (July through September 2023). Like~\citet{guo_computational_2024}, for each post, we retrieved the first two comments and randomized which comment was the "focal comment" that our human annotators were going to label. We provided human annotators with the post and both comments as context, regardless of which comment was the focal post. We randomly assigned each post to two annotators. 

Coders collaboratively went through the first 50 posts in the dataset together to align on their understanding of the sub-labels. After the initial 50 posts, annotators coded in multiple phases. Each phase involved (1) annotating at most 100 posts independently, (2) computing inter-annotator agreement (average Cohen's Kappa \cite{mchugh_interrater_2012} for each annotator pair for each sub-label), (3) analyzing each post where annotators disagreed in order to better-understand and discuss these disagreements, and (4) when applicable, updating definitions in the codebook to reflect a new shared understanding of each label.  

The average Cohen's Kappa per sub-label per round can be found in Appendix A.2. Phase 3 Kappa Scores for sub-labels ultimately included in this research are shown in Table \ref{tab:final_defs}. Some sub-labels yielded particularly low Kappa scores, which could partially be attributed to the relatively low frequency of these sub-labels in our dataset or the subjective nature of their definitions. We chose to simplify the coding task for human annotators and our automated classifier by removing sub-labels from our codebook that appeared less than 1\% of the time (less than four samples) in our dataset. Since we still believe these labels are conceptually relevant to IH/IA, we include descriptions of them in Appendix A. We encourage future research to continue refining the definitions and building larger datasets such that all sub-labels are well represented. After the final round of coding, with infrequent labels removed and label definitions updated, we were able to achieve an average Kappa score of $0.63$ across annotator pairs and sub-labels. A Kappa score in this range represents only ``moderate' agreement''~\cite{mchugh_interrater_2012}. This modest Kappa score highlights the challenging nature of this classification task---even for humans. To ensure a high-quality dataset for future phases of this work and beyond, all three reviewers re-reviewed all 350 posts and updated the assigned sub-labels to reflect group consensus after Kappa scores were calculated. This final dataset represents the ``ground truth'' used later to evaluate our classifier and represents a small gold, standard dataset. This coding process required a conservative estimate of 40 hours of work across the three annotators. We chose to prioritize conceptual alignment of annotators in rating over dataset size due to the difficult nature of this labeling task. Additionally, as we will describe more later, we used a zero-shot LLM-classifier based on conceptual definitions from our codebook and did not require a training dataset for our classifier---using annotated posts instead simply for classifier evaluation---therefore lessening the need for a large ground-truth dataset.

Although the small dataset and low Phase 3 Kappa scores, both discussed more in the Discussion, are limitations of our codebook and coding methodology, they had minimal effects on our downstream classifier and data analysis because of the aforementioned manual review used to construct our gold standard dataset. We include the codebook here in hopes that it will serve as a starting point for researchers wishing to refine and build upon our conceptualizations of IH/IA---and support the development of larger, silver standard datasets.

Table \ref{tab:final_defs} shows the final set of labels, definitions for our codebook, and example posts for each label. We used the fine-grained sub-labels to arrive at a coarse summary categorization for each post (IH/IA/Neutral). Namely, if a post had more IH sub-labels, it was labeled IH. If a post had more IA sub-labels, it was labeled IA. This majority vote classification reflects our belief that a comment can still be overwhelmingly IH, even if it has some IA aspects (and vice versa), but is another limitation of our research. We encourage future research on development of a continuous IH/IA score to reflect the real-life complexity of IH and IA. If a post had no sub-labels, it was labeled neutral. In the rare event that a post had the same number of IA and IH sub-labels, the post was manually split into additional posts such that only IA or IH sub-labels applied to the isolated text to reflect the reality that the equal presence of both IA and IH attributes does not imply neutrality. Once again, a continuous score would likely better capture the nuance of these examples. Our final dataset contained 359 posts.
\begin{table*}[!ht]
    \centering
    \begin{tabular}{@{}p{.8in}cp{2in}p{2in}c@{}c@{}}
    \toprule
        \textbf{Sub-Label} & \textbf{IH/IA} & \textbf{Definition} & \textbf{Example Post} & \textbf{Freq.  } & \textbf{  Kappa} \\
        \midrule
        Acknowledges Personal Beliefs & IH & Affirms individual convictions by speaking openly without contempt and/or uses first-person language to express an opinion or viewpoint without contempt. & \textit{I think a split in the party might be the only possible outcome. The common sense conservatives he talks about are only going to continue to be a dying breed though imo.} & 54 & .87\\
        \addlinespace
        Engages Respectfully with Diverse Perspectives & IH & Directly addresses and thoughtfully responds to differing perspectives in a way that acknowledges their validity or rationale. & \textit{I don't like it, but as a negotiating tactic it can be effective.} & 11 & .38\\
        \addlinespace
        Recognizes limitations in one's own knowledge or beliefs & IH & Acknowledges that one's political knowledge, beliefs, or information sources may be incomplete or subject to bias. & \textit{Nice diagram. I'm not sure what conclusions to draw it, but thank you for posting.} & 8 & 1\\
        \addlinespace
        Seeks out new information & IH & Actively searches for new knowledge and perspectives on political issues or clarification on statements made or poses a non-rhetorical question. & \textit{Did your research include justices who were on the court before the 1900s?} & 18 & .91\\
        \addlinespace
        Polarizing or Tribalistic Language & IA & Characterizes political opponents as inherently evil, less human, or existential threats, creating an "us vs. them" narrative that undermines productive dialogue and fuels division. & \textit{There is one party being led by a criminal trying to break our democracy and another party trying to stop him.} & 21 & .67\\
        \addlinespace
        Condescending Attitude & IA & Overbearing or dismissive behavior that undermines others' perspectives or intellect. & \textit{The main crime is that some people still think the Covid vaccines were successful. Idiots} & 23 & .36\\
        \addlinespace
        Close-minded Absolutism & IA & Using strong, definitive language to express convictions without engaging with or acknowledging diverse perspectives. & \textit{Nobody seems to understand that mere impeachment does ABSOLUTELY NOTHING.... It's NEVER going to happen} & 16 & .22\\
        \bottomrule
    \end{tabular}
    \caption{Sub-Labels, their IH/IA Classification, Definition, an example post from the evaluation dataset, the label frequency in our dataset, and the Phase 3 Kappa Score. Comments have been modified to protect the privacy of original posters.}
    \label{tab:final_defs}
\end{table*}
\begin{table*}[ht]
    \centering
    \begin{tabular}{@{}cp{5.5in}c@{}}
    \toprule
        \textbf{Coarse-Label} & \textbf{Example Post} & \textbf{Freq.} \\
        \midrule
        IA & \textit{We could change every voting rule on the books and literally none of it would matter, moron.} & 55\\
        \addlinespace
        Neutral & \textit{There was an incident in Suffolk County that went unreported that involved election workers leaving ballots unsupervised in the hotel lobby.} & 229 \\
        \addlinespace
        IH & \textit{Honestly, it will be nice to hear some of Rubio's policies. I am backing Haley, but I am also very open to hearing other candidates and how they differ.} & 75\\
        \bottomrule
    \end{tabular}
    \caption{Coarse-Labels, an example post from the evaluation dataset, the label frequency in our dataset, and the Phase 3 Kappa Scores for each sub-label. Comments have been modified to protect the privacy of original posters.}
    \label{tab:placeholder}
\end{table*}
\subsection{Developing the Classifier}
With an operational definition of IH and IA in hand for political discussions on Reddit, we sought to develop an automatic classification method that might enable large-scale annotation of posts using the codebook. This would enable us to analyze large-scale volumes of existing social media data and also form the basis of our intervention (both of which are described in more detail later). Given these broad aims, we focused on developing a coarse-grained classifier to detect IH/IA/Neutral posts---though future work may also further explore the development of fine-grained classifiers that may be more suitable for different tasks, as well as classifiers that consider the broader conversational context, as our human annotators did, when classifying a given post.

\begin{table*}[ht!]
    \centering
    \renewcommand{\arraystretch}{1.2} 
    \begin{tabular}{@{}lcccc@{}}
    \toprule
         \textbf{Classifier} & \textbf{IH} & \textbf{Neutral} & \textbf{IA} & \textbf{Weighted Average} \\
         \midrule
         PerspectiveAPI & $0.53 \pm 0.03$ & $0.82 \pm 0.01$ & $0.23 \pm 0.04$ & $0.67 \pm 0.02$  \\
         GPT-4.1 & $0.66 \pm 0.01$ & $0.80 \pm 0.00$ & $0.56 \pm 0.01$ & $0.74 \pm 0.00$ \\
         \midrule
         Random & $0.18 \pm 0.02$ & $0.62 \pm 0.01$ & $0.14 \pm 0.03$ & $0.47 \pm 0.01$ \\
         Majority Class & $0.00$ & $0.76$ & $0.00$ & $0.47$ \\
         \bottomrule
    \end{tabular}
    \caption{F1 scores and a 95\% confidence interval (where applicable) for: PerspectiveAPI-based classifier (the model uses scikit-learn's default hyperparameters), zero-shot LLM-based classifiers (all labeled data is used for testing), a random baseline classifier, which randomly samples from the distribution of labels in the ground truth data and a majority class classifier (all posts labeled Neutral). We evaluate each classifier 20 times, except the majority classifier which has no randomness}
    \label{tab:classification-models}
\end{table*}

Many of the existing classification methods we found were topic-specific, e.g., for religion \cite{guo_computational_2024} or workplace conflict management \cite{stavropoulos_shadows_2024}; lacked an explicit IA definition \cite{abedin_exploring_2023}; or were adaptations from PerspectiveAPI's bridging dictionary \cite{van_loon_designing_2024}. 
We initially chose to adapt PerspectiveAPI's bridging dictionary to create an IH/IA classifier, but found that doing so led to poor alignment with our human-annotated examples (see Appendix B.1 for additional details, and Table~\ref{tab:classification-models} for a summary of its performance). Given these modest F1 scores (particularly for IA posts) and prior research indicating the limitations of the Perspective API's bridging attributes~\cite{gervais_incivility_2025, rosenblatt_critical_2022}, we turn to Large Language Models (LLMs).

We informally experimented with different zero-shot and few-shot prompting techniques across a number of models. One significant advantage of using an LLM over PerspectiveAPI was that we could apply our codebook directly, rather than abstracting the sub-labels onto the existing PerspectiveAPI attributes. After preliminary explorations, we focused on GPT 4.1, the most promising model and conducted a more extensive evaluation. Table \ref{tab:classification-models} shows evaluation results. The best performing model achieved a weighted average F1-score for the coarse-grained task of 0.74, an improvement over~\citet{guo_computational_2024} (which reported an average F-1 score of 0.7)---promising given the challenging nature of the task but still relatively low. Additionally, since we used a closed-source LLM, we had no measure of confidence for each label. A confidence measure could have been used to develop a continuous IH/IA score. 

Additional details about the LLM process, the exact prompt used, model testing, and how dataset imbalance was handled can be found in Appendix B.2. Error analysis for the classifier can be found in Appendix B.3. Since we qualitatively observed that few-shot prompting led to the classifier's over-reliance on certain example attributes to guide its decisions, our final classifier used a zero-shot prompt. Therefore, as stated earlier, our gold standard dataset was only used to evaluate the effectiveness of the classifier after it was developed (no training/test dataset splitting or cross-validation required). We use the LLM-based classifiers in subsequent analyses given their superior performance across labels.

As shown in Appendix B.3, measurement error for our classifier is almost symmetric. Around 40\% of IH posts are misclassified as Neutral and around 40\% of IA posts are misclassified as Neutral. As we will describe in later sections, all subsequent analysis using this classifier, the Reddit Analysis and the randomized control trial (RCT) analysis, involved labeling posts IA (-1), Neutral (0), and IH (1) and calculating the average score of a user's posts. It is therefore likely that some of the measurement error canceled out during the averaging process. Additionally, since misclassifications erred toward ``Neutral'' rather than the extremes, we believe the associations / effects we observe through the Reddit Analysis and RCT, respectively, may represent conservative estimates of the true underlying trends.

\section{Analyzing Political Conversations on Reddit}
With a definition of IH/IA/Neutral for political discourse in hand and an LLM-based classifier that enables the detection of these attributes at scale, we are now able to explore RQ1: Do pre-existing levels of intellectual humility in online environments influence the expression of intellectual humility in future conversations? To do so, we conduct an observational analysis of existing Reddit data. Importantly, this analysis will not allow us to fully explore this causal question, but enables us to test for associations between discussion environments and the types of posts they eventually house---an important precursor to identifying possible causality.

\subsection{Description of Reddit Data}
We used comment and submission data from ten politics-focused subreddits for a three-month period from March to May 2024. All comments and submissions were collected using Academic Torrents \cite{lo_academic_2016} and filtered to ensure data quality. In pre-processing, we removed deleted and moderated comments (since their content was no longer visible) as well as very short ones (one or two-word replies) and any non-English comments. Comments included existing Reddit data like comment text, author, subreddit, timestamp, and thread identifiers. We additionally derived several fields per comment, namely: an IH/IA/Neutral label generated by our aforementioned LLM classifier; an  ``IH environment'' attribute, which indicates whether a comment was posted in an IH or IA subreddit (an explanation of how this was determined is included in the next section); the rolling mean of IH scores for the $n$ comments before the current one in the same thread to measure a more localized IH environment in the comment thread; and the submission text's topic, derived via BERTopic~\cite{grootendorst2022bertopic}. 
\subsection{Analysis and Results}
\label{sec:redddit-Analysis and results}
We began by using data from March and April 2024 to characterize each subreddit's environment. We applied block sampling to the submission data by randomly selecting 25\% of submissions (and all associated comments) from each day within this period. Based on the IH labels of the sampled comments, where IA, Neutral and IH were coded as -1, 0, and 1 respectively, we calculated the average IH score for each subreddit. Subreddits with an average score greater than zero (r/Anarchism, r/moderatepolitics, r/socialism, r/PoliticalDiscussion) were classified as IH environments, and those less than zero, IA environments (r/Libertarian, r/Foodforthought, r/Conservative, r/politics, r/democrats, r/Republican). For detailed results from this analysis and how we selected these ten politics-focused subreddits see Appendix C.1. \\
\indent After classifying subreddits using data from March and April, we analyzed May 2024 data from these subreddits to explore RQ1. Specifically, we identified all users who commented in at least one IH and one IA subreddit from our set of 10 total subreddits. In total, there were 3,640 such users, and they collectively posted 68,874 comments. To better account for variation in user activity and allow for more reliable comparisons, we further divided the cross-environment users into groups based on their level of engagement. Specifically, we ranked users according to two criteria: (1) their total number of comments across all ten subreddits, and (2) the minimum number of comments they made in either IH or IA environments. For each percentile cutoff (e.g., top 50\%, 25\%, 10\%), we selected users who ranked within that percentile on both criteria. The intersection of these two sets formed the final user groups used in our analysis. For example, the top 10\% group consists of the most active 10\% of cross-environment users according to both criteria: 102 users who collectively contributed 10,837 comments. 

To evaluate whether there is an association between discussion environment and commenting behavior, we first constructed paired user-level IH scores. For each user in the cross-environment group, we calculated the average IH score of their comments made in both IH and IA subreddits, yielding a pair of mean IH scores per user. Although individual IH labels took values -1, 0, and 1, these user-level means vary on a near-continuous scale, so we treated them as continuous outcomes. After observing differences in mean IH scores between IH and IA subreddits to be approximately normal, we ran paired t-tests (and, as a non-parametric robustness check, Wilcoxon signed-rank tests) on these paired IH score means for each user, for each activity subset (top 100\%, 50\%, 25\%, 10\%). Detailed information from those as well as user and comment counts across user groups can be found in Appendix C.2. Results reveal statistically significant differences in levels of IH demonstrated across different types of subreddits (paired t-tests and Wilcoxon tests: $p < 0.001$ for all user subsets, with small but systematic e  ffect sizes, Cohen’s $d \approx 0.16$–$0.38$). To contextualize the difference: approximately one out of every 10 posts that a user posts in an IH environment is more likely to be IH than the the posts they make in an IA environment.

\begin{table}[ht]
    \begin{tabular}{p{1in}|p{3in}}
        & Formulation\\
        Base & $IH_i \sim  Env$ \\
        With Controls & $IH_i \sim Env + C$\\
    \end{tabular}
    \caption{Model specifications. $IH_i$ is an ordinal variable representing the IH classification of comment $i$. $Env$ is an ordinal variable representing the IH classification of the subreddit $i$ was posted in. $C$ represents incrementally introduced control variables, including fixed-effects variables representing author of $i$ to account for poster-level idiosyncrasies, the topic of $i$ to account for associations between topics and posted content and the rolling average IH of recent comments in the same thread as $i$ to capture the IH level of the preceding conversation}
    \label{tab:reddit-modelformulas}
\end{table}

Next, we fit logistic regression models, described in Table~\ref{tab:reddit-modelformulas}, to better understand relationships between environment and posts. We computed these regressions across different user subsets (e.g., comparing results for the top 10\% vs. top 50\% groups) as sensitivity analyses, to examine whether the associations involving key predictors remain consistent across different levels of user activity. Exponentiating model coefficients yielded odds ratios (OR) indicating how the likelihood of expressing intellectual humility increases (OR $>$ 1) or decreases (OR $<$ 1) when in an IH or IA environment. Across our main specifications, the IH environment coefficient remains statistically significant even after applying a conservative Bonferroni correction across the four activity groups and five model specifications (20 tests in total), suggesting that the observed associations are unlikely to be driven purely by chance.

Results from these regressions further highlight the ways in which environment is associated with levels of IH/IA demonstrated in comments. Across all user groups and model specifications, the subreddit IH environment score consistently emerged as a statistically significant and practically-meaningful predictor of the comment's IH score. For example, in the Top 50\% group, the coefficient for the IH environment score was $\beta \approx 0.56$ when no other control variables are included. After controlling for author identity, topic, and the tone of preceding thread discourse, the coefficient remained significant (though decreased to $\beta \approx 0.38$). This corresponds to an odds ratio of $\approx 1.46$, indicating that comments made in IH environments had roughly 46\% higher odds of being labeled IH than comments made in IA environments, even after adjusting for various individual and contextual factors. We observed a similar pattern for the rolling mean IH score within threads, which captures the tone of recent comments within the same thread (i.e., the ``microenvironment''). Appendix C.3 contains the full regression specification and regression tables describing these and other results. Our results highlight a relationship between environmental and individual discourse patterns.


There are three important caveats with this analysis. One relates to the classifier we use. Although the GPT-4.1-based classifier performs well on our gold-standard evaluation set, it is still imperfect. Any residual misclassifications likely introduce measurement error; however, as described earlier, we believe these misclassifications are likely to conservatively understate the true strength of the estimated associations. 
Second, the pseudo $R^2$ values of our models are lower than one might hope, as are the regression coefficients, suggesting a weak relationship between model inputs and future expressions of IH/IA. This makes sense, however, because in practice, whether a single Reddit comment expresses IH depends on many factors that we do not explicitly model, so environment, author, and topic can only explain a limited portion of the variance. Our goal here is to test whether the IH environment is systematically associated with comment-level IH once we control for observable factors. From this perspective, the most important quantities are the direction and robustness of the IH environment coefficients, which remain statistically significant across user groups and model specifications, even after correcting for multiple comparisons.
Third, the results suggest an associational, and not causal, relationship between the average IH/IA level of a subreddit and the IH/IA levels of comments subsequently posted in it. Controlling for user, topic, and thread helps rule out potential confounders and makes the observed associations more interpretable, but there remain important unobserved confounders that these variables do not account for---for example, users’ personalities, momentary mood, fine-grained subtopics, conversational history, and the mindsets and perceived norms they bring to different environments. For instance, the same Reddit user might naturally behave differently in different environments (e.g., because of the mindset they adopt there and the norms they perceive). These confounders could both impact users' propensities to participate in particular subreddits and affect the nature of their posts, and therefore, prevent causal interpretations in this secondary data analysis. To explore causality, we turn to randomized experiments.

\section{Lab-Based Randomized Experiment}
The Reddit analysis identified the potential presence of an intellectual humility contagion effect on Reddit---where the marginal post in an intellectually humble or arrogant environment is likely to be intellectually humble or arrogant, respectively. Crucially, however, ``contagion'' suggests a causal effect, while our Reddit analysis only revealed associations at best. To investigate possible causal effects and how, therefore, posting behavior might be modified away from impacts of discussion environments (specifically, in cases where environments demonstrate high levels of IA), we designed and engineered a novel social media testing environment to support a lab-based randomized experiment. In particular, the experiment explores the extent to which a nudge that prompts a participant to reflect on ways to make their posts demonstrate more intellectual humility changes what is posted (regardless of how IH/IA the discussion environment already is). The underlying hypothesis here is that in face-to-face conversations, we are often privy to social cues (facial expressions, body language, etc.) that might steer us toward more prosocial / less incendiary discourse. While such cues are lacking in online spaces, nudges like this one---that prompt reflection and offer guidance on how to change posts---may help. 

\subsubsection{Pre-Survey and Consent} We conduct a pre-registered (\url{https://aspredicted.org/q7g4-dvk9.pdf}) randomized control trial on the research platform Prolific to understand the effectiveness of this intervention across different discussion environments. All participants first completed a pre-survey to collect information on their self-reported intellectual humility and self-reported interest and stance across three topics: Abortion, Climate Change, and Immigration. Survey questions can be found in Appendix D.1. We also collected informed consent from all participants through a form shown in Appendix D.2. Visualizations of the social media test environment can be found in Appendix D.3.

\subsubsection{Randomization}
Participants were randomly assigned into one of several treatment or control conditions across two orthogonal experiments - 
Social Cue" and "Environment". In total, there were six possible experimental groups participants could be placed in: 2 ("Social Cue" received or not) x 3 ("Environment" - IA, Neutral, IH).

\subsubsection{Social Media Task and Environment Intervention}
All participants were asked to engage with at least two different Reddit-style posts. The topic of the posts and subsequent conversation was determined by the participant's self-reported interest and stance from the pre-survey. The pre-seeded posts were deliberately designed to depict viewpoints that were likely to oppose the participants' stance. 

Participants were asked to submit an intended comment in response to the pre-seeded posts. As described below, these intended comments were handled differently depending on which ``Social Cue" group participants were placed in. 

Once intended comments were posted, one of 10 AI dialogue agents prompt-engineered to represent a certain persona (conditioned on a particular gender identity, age, and occupation) was chosen to respond. Prompts for these personas can be found in Appendix D.4. For each participant, the demeanor of the AI dialogue agents, which could be IH, IA, or Neutral, depended on which of the "Environment" groups, IH, IA, or Neutral, participants were randomly placed in. The prompts used to establish the demeanor of these bots can be found in Appendix D.6. 

Once a bot had responded, participants then had an opportunity to respond, and so on and so forth. Participants were required to make at least two comments on each post. 

\subsubsection{Social Cue Intervention}
For participants randomized into the treatment arm of the first experiment (``Social Cue"), we used the GPT-4.1 classifier described and evaluated earlier to determine (in real-time) if their intended comment was IH, IA or Neutral. If participants submitted an IH comment, that comment was automatically posted to the social media feed. If participants submitted an IA or Neutral comment, their comment was not posted to the social media environment feed. Instead, participants received personalized feedback generated by a separate GPT-4.1 prompt with suggestions for how they might make their comment more IH. We flagged comments not reflecting IH, instead of simply those demonstrating IA, in order to advance more virtuous discourse through proactive demonstrations of IH instead of simply an absence of IA. After receiving this feedback, participants could decide to post their original comment or modify it by posting the LLM-generated suggestion or some derivation of it. The prompt used to provide this feedback can be found in Appendix D.5. 

Figure~\ref{fig:intervention-flow} in Appendix D demonstrates the experiment flow. We retained both intended and posted comments and the GPT 4.1 classifications generated during the experiment.  

For participants randomized into the control arm of the first experiment (``Social Cue"), intended comments were posted automatically with no feedback provided to the participant, regardless of if the comments were IH, Neutral, or IA. We retained all posted comments and used our GPT 4.1 classifier to classify them after the experiment concluded.

\subsubsection{Post-Survey} After leaving at least two comments on each of the two parent posts, participants completed a post-survey collecting demographic information and, again, self-reported intellectual humility.

\subsubsection{Compensation}
The wage for each task was set to \$2.50 (pro-rated from an hourly wage of \$15/hour, based on an estimate of 10 minutes spent per task inferred through a test run of the experiment). After running the experiment, the average compensation was approximately \$12/hour (participants ended up taking longer on the full study than the test). 



\subsubsection{Data Collection \& Participant Information} All data were collected and stored in a MongoDB database, through a web application with a front-end built in ReactJS and a backend in Flask. All responses were anonymous, and participants were asked not to share identifying information.

We enrolled 513 American, self-identified social media users, with political ideologies quota-sampled to reflect the prevalence of ideological backgrounds across the US. Of these, 412 successfully completed the task; however, only 355 passed all of the attention checks, finished both the pre- and post-survey, and left the requisite number of comments in our social media testing environment. We compensated the 412 participants but use the set of 355 in our analysis.

\subsection{Results}
\begin{table}[t]
\centering
\begin{tabular}{p{2cm} p{1.8cm} p{1.8cm} p{1cm}}
\toprule
Environment & Social Cue (Treated) & Social Cue Control & Total \\
\midrule
IH      & 56 (53) & 53 & 109 \\
Neutral & 63 (63) & 62 & 125 \\
IA      & 56 (54) & 65 & 121 \\
\midrule
\textbf{Total} & \textbf{175 (170)} & \textbf{180} & \textbf{355} \\
\bottomrule
\end{tabular}
    \caption{Number of participants across two experiments; numbers in parentheses indicate those whose posts were flagged at least once for feedback.}
    \label{tab:participantbreakdown}
\end{table}

\begin{figure}[t]
    \centering
    \includegraphics[width=\linewidth]{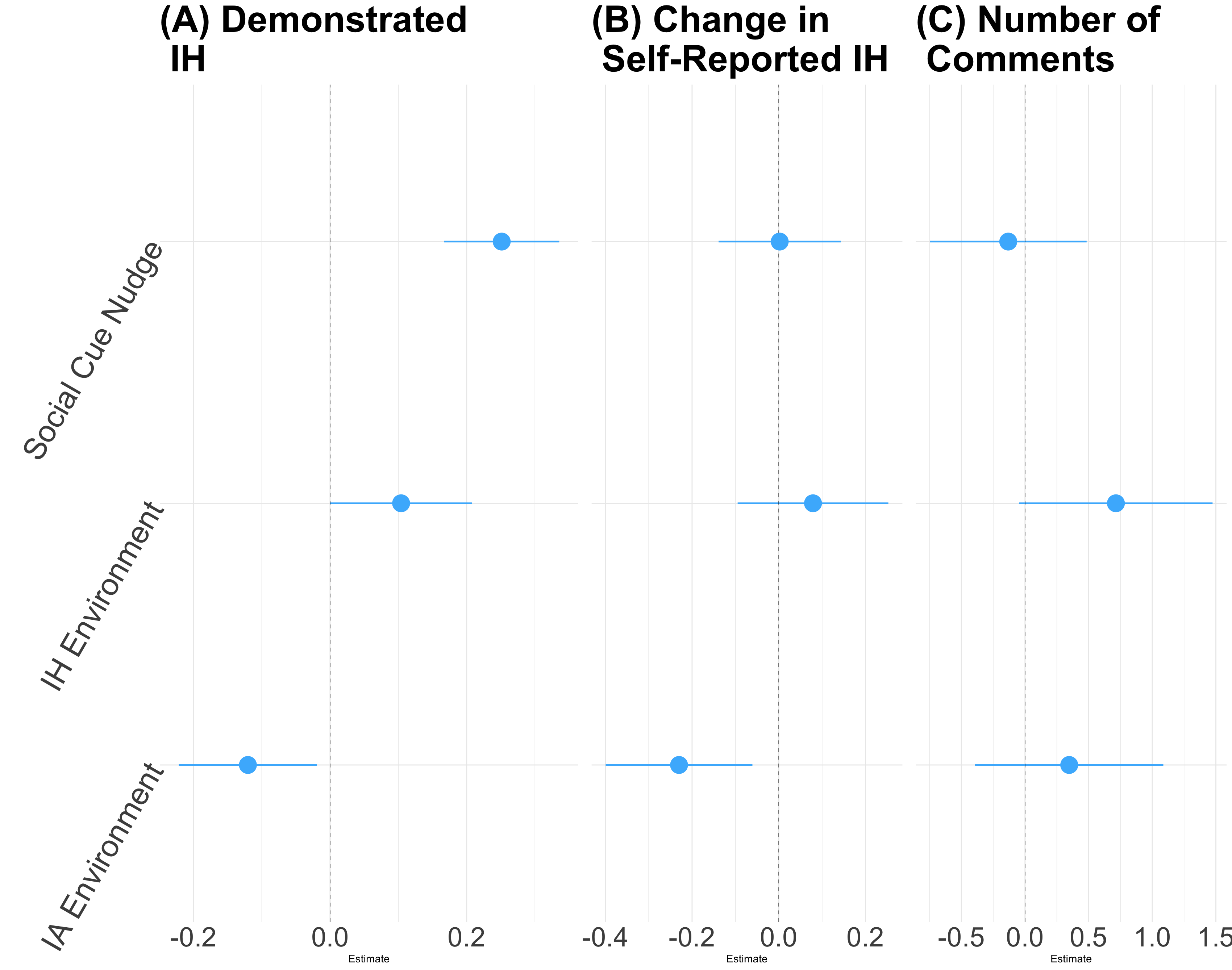}
    \caption{The coefficients for the ``Base Model'' shown in Table \ref{tab:modelformulas} with three outcomes of interest. Error bars represent 95\% confidence intervals.}
    \label{fig:intent-to-treat-main-effects}
\end{figure}

\begin{figure}[t]
    \centering
\includegraphics[width=\linewidth]{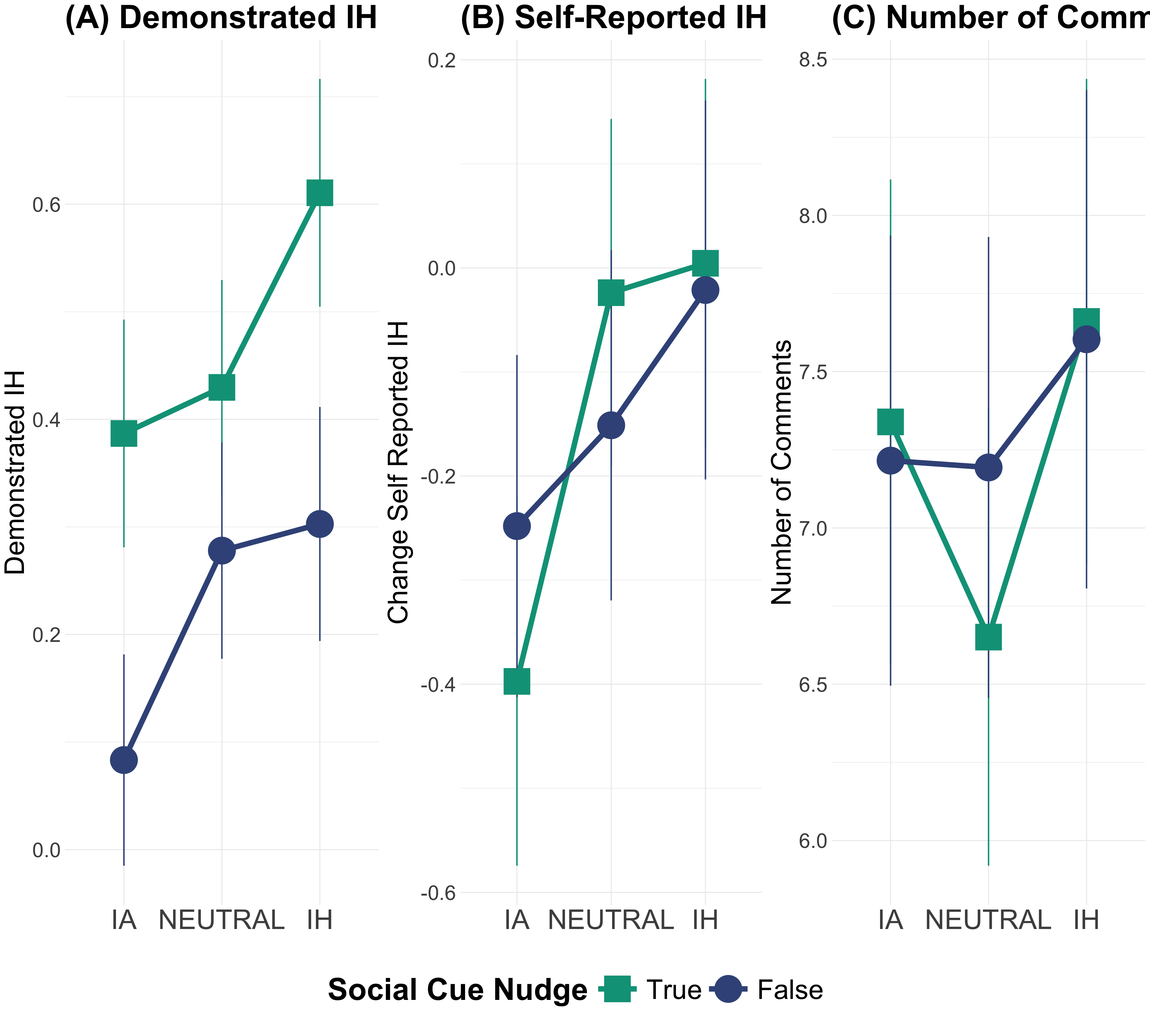}
    \caption{Interactions between the effects of the treatments for the ``Interaction Model'' shown in Table\ref{tab:modelformulas}. Plotted values are predicted outcomes of interest from each model and slopes indicate interaction effects. Error bars represent 95\% confidence intervals.}
    \label{fig:intent-to-treat-marginal-effects}
\end{figure}

Table \ref{tab:participantbreakdown} shows participants per experimental group. We conducted a conservative intent-to-treat analysis to examine the impact of our two lab-based experiments on three outcomes of interest: demonstrated intellectual humility, change in self-reported intellectual humility, and number of comments per participant (this last outcome was not pre-registered, so we report it here as an additional exploratory analysis). To calculate demonstrated IH, comments were labeled and IH scores were calculated using the same method as described in the Reddit Analysis. We refer to these IH scores as ``demonstrated'' IH scores to avoid confusion with ``self-reported'' IH scores. The change in self-reported intellectual humility for participant $p$ was computed as the difference between the $p$'s average response to the 10-point Likert-style intellectual humility questions in the post-survey and the average response to the same questions in the pre-survey. Number of comments was designed to represent the overall conversation length as a proxy for engagement. Although our social cue intervention may enhance discourse quality, social media interventions that decrease platform engagement more generally are unlikely to gain traction outside of academic settings, as they do not align with social media platforms' commercial incentives. \\
\indent Our first model (``Base Model'') represents the most conservative model, containing only dummy variables for each treatment. This model presents the causal effect of both treatments on our three different dependent variables (we include analyses that include participant covariates, as described in our pre-registration, in Appendix E.2). Our second model (``Interaction Model'') explores how the effects of the ``Social Cue'' treatment vary across different Environments. Table~\ref{tab:modelformulas} specifies the models. \\
\indent Figure \ref{fig:intent-to-treat-main-effects} shows the coefficients from the ``Base'' model for the three outcomes of interest, and Figure~\ref{fig:intent-to-treat-marginal-effects} shows results from the ``Interaction'' model. To interpret the magnitude of coefficients for demonstrated IH: a participant's demonstrated IH score can be thought of as the average of the scores of their posts. Therefore, an effect of 0.25 from receiving the ``Social Cue'' treatment indicates a 25\% increase in the IH demonstrated by a participant's post after being exposed to this treatment. Put another way: on average, one out of every four of a treated participant's IA posts is now likely to be neutral; or one out of every four of their neutral posts is now likely to be IH; or one out of every 8 of their IA posts is now likely to be IH. For the Self-Reported IH outcome, the magnitude of the coefficient represents the average change in their likert-scale response to questions assessing levels of IH. Since we used a discrete scale from 1-10 on the survey, our IA coefficient of -.23 means that participants in the IA group self-reported a one point lower score on 1-2 of eight questions assessing IH following treatment, relative to before. These questions can be found in Appendix D.1.
\begin{table}[ht]
    \centering
    \begin{tabular}{p{.55in}|p{3in}}
        &Formulation\\
        Base & $Y_i = \beta_0 + \beta_1 \cdot C + \beta_2 \cdot E$ \\
        Interaction & $Y_i = \beta_0 + \beta_1 \cdot C  + \beta_2 \cdot E  + \beta_3 \cdot (C \cdot E$)\\
    \end{tabular}
    \caption{Model specifications. $Y_i$ represents one the three outcomes of interest - demonstrated IH, self-report IH, or number of comments; $C$ is a binary variable indicating whether or not a participant was assigned to the ``Social Cues'' treatment group; $E$ represents which environment a participant was assigned to (IA, Neutral, or IH).}
    \label{tab:modelformulas}
\end{table}

Analyzing the results of these models confirms some of the findings from our Reddit data analysis---most notably, the potential for an IA contagion effect. Users randomized into an IA Environment demonstrated, on average, more IA behavior compared to users in the ``Neutral Environment''. Users in the 'IH Environment' demonstrated, on average, more IH behavior compared to users in the ``Neutral Environment''. Additionally, The ``Social Cue'' treatment had a significant, positive effect on demonstrated IH. As shown in Figure \ref{fig:intent-to-treat-marginal-effects}, the effectiveness of the nudge on promoted demonstrated IH varied based on environment---with a larger effect in the extreme (IH, IA) environments. Full regression tables for these models can be found in Appendix E.1.   

The ``Social Cue'' Nudge had no impact on self-reported IH, perhaps for several reasons. For one, self-reported IH may be a lagging indicator: Even though behavior changes, people must reflect before they identify the change in themselves, if at all. Alternatively, participants may not view their behavior towards bots in our mock social media environment as relevant to their sense of IH. Their ``real-world'' interactions may matter more to them. Interestingly, there was a small but statistically significant negative impact of IA environments on self-reported IH in the ``Base Model''. This may suggest that people are hesitant to update their sense of self-reported IH based on nudge-style feedback but are subconsciously impacted by the behavior of those around them---and perhaps view it as a reflection of, or strong influence on, their own underlying demeanor. However, our study was not designed to explicitly explore these dynamics.  Additionally, the self-reported IH results were small and inconsistent (which are particularly clear when looking at the interaction effects depicted in Figure~\ref{fig:intent-to-treat-marginal-effects}). Future research should continue to explore self-reported IH as a construct and as a target of intervention. 
Our self-reported IH results suggest that future interventions need not necessarily change individual's perceptions of their own IH, which is challenging and likely requires a longer time horizon. To positively impact conversation dynamics, it may be sufficient to enhance demonstrated IH---a more malleable intermediate goal.

With respect to engagement, we found no significant relationship between either the ``Environment'' or ``Social Cue'' intervention and either engagement metric. In fact, users across all treatment groups posted more than 6 comments (on average), even though they were only required to make four posts in total (two in each of the discussions they were embedded in). These findings echo prior work~\cite{stroud_intellectual_2025} and suggest that fostering more IH online need not come at the expense of engagement (and therefore, commercial) objectives. This is supplemented by the fact that often times, users may even actively avoid IA environments: indeed, many participants in the IA environment treatment concluded their participation because they believed the conversation environment had become too toxic (this was an anticipated risk submitted with the IRB application that was subsequently approved and enabled this study). Of course, given the sandboxed nature of our lab-based experiment, these findings should be further evaluated \textit{in situ} through field experiments in live discussion environments. 

Appendix E.3 reports the additional pre-registered specifications of these models---i.e., those including participant covariates. These covariates include the specific topics/stances of a given participant's conversation and the (baseline) level of IH self-reported in the pre-survey. One interesting observation is that baseline levels of self-reported IH are predictive of more demonstrated IH post treatment.



\section{Discussion}
Recall our original research questions from the introduction: in regards to RQ1, we find suggestive (associational) evidence through our secondary Reddit data analysis and confirmatory (causal) evidence through our lab-based randomized experiment that pre-existing levels of IH and IA can influence the extent to which these attributes, respectively, are demonstrated in future posts. This finding extends work in related domains like toxicity~\cite{saveski2021} to the domain of Intellectual Humility and advances related findings from recent work~\cite{stroud_intellectual_2025}. In regards to RQ2, we find that our ``Social Cue'' intervention successfully enhances levels of IH demonstrated across posts, on average, across different conversation environments (those that demonstrate more IH or IA) and politically-charged topics. These findings build on promising results from other studies that show how dynamic suggestions to users at the time of making posts can enhance discussion quality~\cite{katsaros2021,Argyle2023_democratic_discourse} and show how such interventions can be robust to different conversational dynamics.

While our study offers several contributions in the measurement and study of intellectual humility in online public discourse, it also has a number of limitations. For one, given the resource-intensive nature of qualitatively annotating posts for IH/IA, our gold-standard dataset is relatively small at 359 posts. Low inter-rater reliability (Kappa scores) for the IH/IA sub-labels is an additional limitation. Since we manually reviewed all labels before using our gold standard dataset in downstream tasks, we believe these low Kappa scores have minimal spillover effects on our classifier, Reddit Analysis, and controlled experiment. Still, we did use the sub-label definitions in our LLM prompt, though the target of the classification task was ultimately the coarse IH/IA/Neutral labeling, mitigating the impact that lower sub-label agreement might have on our ensuing analyses. The high variation in Kappa scores for our sub-labels indicates that the sub-label approach to IH/IA identification is promising, especially for more objective concepts--such as ``Seeks out new information'', but requires significant refinement for more subjective concepts--such as ``Close-minded Absolutism''. In general, our classifier achieves a weighted average F1 score across classes of 0.74---a large improvement over a naive random baseline (0.47), and improvement over related prior work in the Religion domain~\cite{guo_computational_2024} (0.7), but still leaving much room for improvement. Still, as described earlier, we believe these errors yield conservative findings and likely understate conclusions from the Reddit and experimental analyses.  

Another limitation is that our randomized experiment is conducted in a sandboxed environment, raising questions about the generalizability of findings. While we sought to retain as many key characteristics of social media as possible---e.g., sorting participants into topic groups they indicated an interest in; having multiple bot personas to simulate online group discourse; and mimicking the vertical scroll design of online platforms---our mock social media environment is missing key aspects of true social media. Most notably, participants knew they were interacting with bots, which may have impacted the nature of their discourse. Additionally, participants were in overwhelmingly IA/Neutral/IH environments when different social media environments may display a more heterogeneous mixing of IA/Neutral/IH content. It is also not clear what long-term effects, if any, the intervention had on participants. Similar findings from the Reddit analysis should assuage some concerns about generalizability, but replications of this study and/or extensions to field deployments are important to further contextualize and enhance the findings. 

Finally, our hard classification of posts as IH/IA/Neutral limits our ability to conduct more nuanced analyses that a softer-classification scheme (e.g., computing the degree to which a given post reflects IH/IA/Neutrality) might enable. Future work may explore alternative classification schemes.

Beyond methodological limitations, our study also raises philosophical questions about the value of promoting IH in discussion spaces. For example, our lab-based randomized experiment found that participants with higher levels of baseline self-reported IH are more likely to demonstrate increased IH in their posts (irrespective of which treatment conditions they are enrolled in for each experiment). The existence of a potential IH contagion effect suggests there may be benefit to this overall increase in IH, thought more research is certainly needed to understand if the potential contagion effect exists and, if it does, whether there are eventually diminishing marginal returns from IH behavior. Despite the many direct and residual benefits of more IH in discussion spaces described in the introduction and throughout the paper, critics may argue that making already IH users more IH could actually widen chasms and create more opportunity for less IA, and perhaps more extreme, voices to dominate. This could be especially undesirable in online spaces, where content is curated through algorithms designed to promote engagement---not intellectual humility. This concern is entrenched in a ``zero sum'' view of political discourse, where the concessions or accommodations made by one group may be perceived as a weakness, and subsequently indirectly fortify, the arguments of another. While we do not subscribe to this view, it is important to keep in mind and investigate such potential effects through future research. 

Indeed, we urge researchers (including ourselves) to approach the future study of intellectual humility in online discourse with intellectual humility: a genuine curiosity and desire to better understand the impacts that fostering IH might have on individuals and discussion dynamics, without an attachment to particular outcomes. We invite researchers from a wide range of disciplines to envision new studies that build on both the contributions and limitations of this paper in order to explore the philosophical, empirical, and other dimensions of this important topic. 
\section{Ethical Considerations}
In addition to receiving IRB approval, we made it clear to participants in the randomized experiment they could leave at any point if they experienced discomfort. To protect the privacy of users in the Reddit analysis, we did not conduct any user specific analysis. We also modified all Reddit posts included in this paper to protect user privacy. 

\section{Acknowledgments}
We are grateful to the Templeton Foundation and Georgia State University for supporting this work.

\bibliography{aaai2026}

\clearpage
\subsection{Paper Checklist}

\begin{enumerate}

\item For most authors...
\begin{enumerate}
    \item  Would answering this research question advance science without violating social contracts, such as violating privacy norms, perpetuating unfair profiling, exacerbating the socio-economic divide, or implying disrespect to societies or cultures?
    \answerYes{Yes}
  \item Do your main claims in the abstract and introduction accurately reflect the paper's contributions and scope?
    \answerYes{Yes}
   \item Do you clarify how the proposed methodological approach is appropriate for the claims made? 
    \answerYes{Yes}
   \item Do you clarify what are possible artifacts in the data used, given population-specific distributions?
    \answerYes{Yes}
  \item Did you describe the limitations of your work?
    \answerYes{Yes}
  \item Did you discuss any potential negative societal impacts of your work?
    \answerYes{Yes}
      \item Did you discuss any potential misuse of your work?
    \answerYes{Yes}
    \item Did you describe steps taken to prevent or mitigate potential negative outcomes of the research, such as data and model documentation, data anonymization, responsible release, access control, and the reproducibility of findings?
    \answerYes{Yes}
  \item Have you read the ethics review guidelines and ensured that your paper conforms to them?
    \answerYes{Yes}
\end{enumerate}

\item Additionally, if your study involves hypotheses testing...
\begin{enumerate}
  \item Did you clearly state the assumptions underlying all theoretical results?
    \answerYes{Yes}
  \item Have you provided justifications for all theoretical results?
    \answerYes{Yes}
  \item Did you discuss competing hypotheses or theories that might challenge or complement your theoretical results?
    \answerYes{Yes}
  \item Have you considered alternative mechanisms or explanations that might account for the same outcomes observed in your study?
    \answerNo{No}
  \item Did you address potential biases or limitations in your theoretical framework?
    \answerYes{Yes}
  \item Have you related your theoretical results to the existing literature in social science?
    \answerYes{Yes}
  \item Did you discuss the implications of your theoretical results for policy, practice, or further research in the social science domain?
    \answerYes{Yes}
\end{enumerate}

\item Additionally, if you are including theoretical proofs...
\begin{enumerate}
  \item Did you state the full set of assumptions of all theoretical results?
    \answerNA{N/A}
	\item Did you include complete proofs of all theoretical results?
    \answerNA{N/A}
\end{enumerate}

\item Additionally, if you ran machine learning experiments...
\begin{enumerate}
  \item Did you include the code, data, and instructions needed to reproduce the main experimental results (either in the supplemental material or as a URL)?
    \answerNo{No — these will be finalized following anonymous peer review.}
  \item Did you specify all the training details (e.g., data splits, hyperparameters, how they were chosen)?
    \answerYes{Yes}
     \item Did you report error bars (e.g., with respect to the random seed after running experiments multiple times)?
    \answerYes{Yes}
	\item Did you include the total amount of compute and the type of resources used (e.g., type of GPUs, internal cluster, or cloud provider)?
    \answerNA{N/A}
     \item Do you justify how the proposed evaluation is sufficient and appropriate to the claims made? 
    \answerYes{Yes}
     \item Do you discuss what is ``the cost`` of misclassification and fault (in)tolerance?
    \answerNo{No}
  
\end{enumerate}

\item Additionally, if you are using existing assets (e.g., code, data, models) or curating/releasing new assets...
\begin{enumerate}
  \item If your work uses existing assets, did you cite the creators?
    \answerYes{Yes}
  \item Did you mention the license of the assets?
    \answerNA{N/A}
  \item Did you include any new assets in the supplemental material or as a URL?
    \answerYes{Yes}
  \item Did you discuss whether and how consent was obtained from people whose data you're using/curating?
    \answerNA{N/A}
  \item Did you discuss whether the data you are using/curating contains personally identifiable information or offensive content?
    \answerYes{Yes}
\item If you are curating or releasing new datasets, did you discuss how you intend to make your datasets FAIR (see \citet{fair})?
\answerNo{No — we will adhere to these standards when preparing the final release of post annotations and describe this in the paper.}
\item If you are curating or releasing new datasets, did you create a Datasheet for the Dataset (see \citet{gebru2021datasheets})? 
\answerNo{No - we will do this after anonymous peer review, once we release the data.}
\end{enumerate}

\item Additionally, if you used crowdsourcing or conducted research with human subjects...
\begin{enumerate}
  \item Did you include the full text of instructions given to participants and screenshots?
    \answerYes{Yes}
  \item Did you describe any potential participant risks, with mentions of Institutional Review Board (IRB) approvals?
    \answerYes{Yes}
  \item Did you include the estimated hourly wage paid to participants and the total amount spent on participant compensation?
    \answerYes{Yes}
   \item Did you discuss how data is stored, shared, and deidentified?
   \answerYes{Yes}
\end{enumerate}

\end{enumerate}

\appendix
\maketitle

\clearpage

\section{Appendix A: IH Codebook Development}
Table \ref{tab:def_change_table} describes the modifications made from the IH codebook for religion presented by \citeauthor{guo_computational_2024}

\onecolumn
\begin{sidewaystable}[p]
\centering
\caption{Summary of re-definitions from the religious codebook to the politics codebook, where `---' indicates that label was removed and * indicates the label was omitted from final analysis in this research}
\label{tab:def_change_table}
\vspace{0.5em}
\renewcommand{\arraystretch}{1.4}
\normalsize
\hspace*{\fill}
\begin{tabular}{>{\raggedright\arraybackslash}p{2.2cm}>{\raggedright\arraybackslash}p{2.2cm}>{\raggedright\arraybackslash}p{3.5cm}>{\raggedright\arraybackslash}p{3.5cm}>{\raggedright\arraybackslash}p{8cm}}
        \hline
        \textbf{Religious Label} & \textbf{Politics Label} & \textbf{Religious Definition} & \textbf{Politics Definition} & \textbf{Justification for Change} \\
        \hline
        Acknowledges Personal Beliefs & Acknowledges Personal Beliefs & Affirms individual convictions with the recognition that they are personal perspectives, open to interpretation. & Affirms individual convictions by speaking openly without contempt and/or uses first-person language to express an opinion or viewpoint without contempt. & The name of this label was retained from \citeauthor{guo_computational_2024}, but the definition for this label was modified to explicitly include "and/or uses first-person language to express an opinion or viewpoint without contempt" in an attempt to make labeling more objective. \\
        \hline
        Respects Diverse Perspectives & Engages Respectfully with Diverse Perspectives & Acknowledges a different perspective in one's statement, and gives it consideration and value. & Directly addresses and thoughtfully responds to differing perspectives in a way that acknowledges their validity or rationale. & This wording change highlights the importance of actively engaging with different viewpoints in a meaningful and thoughtful manner, rather than just acknowledging them. This is particularly important for coders to keep in mind as research has shown that, although some perspective-taking can have positive outcomes \cite{todd_perspective-taking_2014}, not all perspective-taking is productive as it can lead to harmful feelings towards other groups \cite{tarrant_social_2012, todd_perspective_2013}. \\
        \hline
        Recognizes Limitations in one's own knowledge or beliefs & Recognizes Limitations in one's own knowledge or beliefs & Understands that personal religious knowledge or beliefs might not be complete or fully accurate. & Acknowledges that one's political knowledge, beliefs, or information sources may be incomplete or subject to bias. & This modification in definition is to apply more specific language for the politics domain. \\
        \hline
        Seeks Out New Information & Seeks Out New Information & Actively looks for new knowledge and perspectives about different religions or clarification on statements made. & Actively searches for new knowledge and perspectives on political issues or clarification on statements made and/or posing non-rhetorical questions. & The name of this label was retained from \citeauthor{guo_computational_2024}, but the definition for this label was modified to explicitly include "and/or posing non-rhetorical questions" in an attempt to make labeling more objective. \\
        \hline
\end{tabular}
\hspace*{\fill}
\end{sidewaystable}
\clearpage

\begin{sidewaystable}[p]
\centering
\renewcommand{\arraystretch}{1.4}
\normalsize
\hspace*{\fill}
\begin{tabular}{>{\raggedright\arraybackslash}p{2.2cm}>{\raggedright\arraybackslash}p{2.2cm}>{\raggedright\arraybackslash}p{3.5cm}>{\raggedright\arraybackslash}p{3.5cm}>{\raggedright\arraybackslash}p{8cm}}
        \hline
        \textbf{Religious Label} & \textbf{Politics Label} & \textbf{Religious Definition} & \textbf{Politics Definition} & \textbf{Justification for Change} \\
        \hline
        Mindful of Others' Feelings & Displays Empathy* & Considers how religious discussions or actions might affect others emotionally. & Demonstrates an understanding of and sensitivity to other people in the argument's emotional experiences. & This change represents a shift toward a more precise concept in political discourse - emphasizing clear, observable behaviors that reflect genuine emotional understanding. Empathy in political contexts involves actively recognizing and addressing how political arguments and positions may affect individuals emotionally, which can encourage individuals to reflect on political issues at a deeper level and elicit greater feelings of concern towards those with opposing viewpoints \cite{muradova_reflective_2022} \\
        \hline
        Reconsiders beliefs when presented with new evidence & Reconsiders beliefs when presented with new evidence* & Willingness to rethink religious beliefs when faced with new information that challenges them. & Demonstrates a willingness to rethink political beliefs when credible, new information challenges previous assumptions within the thread. & This modification in definition is to apply more specific language for the politics domain and setting a higher bar for evidence to account for the prevalence of misinformation in political discourse \cite{jerit_political_2020}. \\
        \hline
        Displays Absolutist Language & Close-Minded Absolutism & Uses rigid language implying there's only one absolute truth in religion. & Using strong, definitive language, including but not limited to phrases like never or always, to express convictions without engaging with or acknowledging diverse perspectives. & This change recognizes that strong convictions are not inherently problematic in political discourse; what signals intellectual arrogance is the outright dismissal of other viewpoints. This shift reflects research by \citeauthor{hannon_is_2024} which highlights the potential tension between holding strong political convictions and the willingness to engage with opposing viewpoints. \\
        \hline
        --- & Polarizing or Tribalistic Language & --- & Characterizes opposing groups as inherently evil, less human, or existential threats, creating an ``us vs. them'' narrative that undermines productive dialogue and fuels division. & This label highlights language that demonizes or dehumanizes opponents, fostering an adversarial mindset that obstructs constructive dialogue. It has been shown that even in non-political settings, political polarization leads individuals to discriminate against opposing partisans \cite{iyengar_fear_2015}. Intellectual humility has been linked to less affective polarization \cite{bowes_how_2024}. We are extending this idea by saying that intentional polarization, which frames discussions as battles between two groups rather than an opportunity to grow, is a component of intellectual arrogance. \\
        \hline
\end{tabular}
\hspace*{\fill}
\end{sidewaystable}
\clearpage

\begin{sidewaystable}[p]
\centering
\renewcommand{\arraystretch}{1.4}
\normalsize
\hspace*{\fill}
\begin{tabular}{>{\raggedright\arraybackslash}p{2.2cm}>{\raggedright\arraybackslash}p{2.2cm}>{\raggedright\arraybackslash}p{3.5cm}>{\raggedright\arraybackslash}p{3.5cm}>{\raggedright\arraybackslash}p{8cm}}
        \hline
        \textbf{Religious Label} & \textbf{Politics Label} & \textbf{Religious Definition} & \textbf{Politics Definition} & \textbf{Justification for Change} \\
        \hline
        --- & Overinflated Expertise* & --- & Exaggerates one's own expertise or experience to make overly definitive or generalized assertions. & This label captures instances where individuals misrepresent the depth of their own knowledge, either by presenting themselves as experts without providing sufficient credentials or by drawing sweeping, close-minded conclusions from personal experience. While there is value to sharing personal anecdotes, \citeauthor{light_role_2020} suggest that knowledge mis-calibration is a significant barrier to intellectual humility. \\
        \hline
        Condescending Attitude & Condescending Attitude & Arrogant or dismissive behavior that undermines others' perspectives or intellect & Overbearing or dismissive behavior that undermines others' perspectives or intellect. & This label and definition is retained from \citeauthor{guo_computational_2024}. \\
        \hline
        Ad Hominem & Ad Hominem* & The argument attacks the person making the argument instead of addressing the argument itself. & The argument attacks the person making the argument instead of addressing the argument itself. & This label and definition is retained from \citeauthor{guo_computational_2024}. \\
        \hline
        Displays Prejudice & Displays Prejudice* & Unfair opinions or judgments about someone or a group without proper understanding or reason, often based on factors like race, religion, or gender. & Unfair opinions or judgments about someone or a group without proper understanding or reason, often based on factors like race, religion, or gender. & This label and definition is retained from \citeauthor{guo_computational_2024}. \\
        \hline
\end{tabular}
\hspace*{\fill}
\end{sidewaystable}
\clearpage

\begin{sidewaystable}[p]
\centering
\renewcommand{\arraystretch}{1.4}
\normalsize
\hspace*{\fill}
\begin{tabular}{>{\raggedright\arraybackslash}p{2.2cm}>{\raggedright\arraybackslash}p{2.2cm}>{\raggedright\arraybackslash}p{3.5cm}>{\raggedright\arraybackslash}p{3.5cm}>{\raggedright\arraybackslash}p{8cm}}
        \hline
        \textbf{Religious Label} & \textbf{Politics Label} & \textbf{Religious Definition} & \textbf{Politics Definition} & \textbf{Justification for Change} \\
        \hline
        Closed to Diverse Perspectives & --- & Unwillingness to consider, engage, or accept viewpoints different from one's own in religion. & --- & Human coders and language models struggled to reliably apply this label because of its ambiguous definition. \citeauthor{roberts_intellectual_2007} suggest that intellectual arrogance is a refusal to consider other perspectives and break this high-level concept into two clearer characterizations: overconfidence in one's own beliefs and a dismissive attitude toward opposing views. We embrace this view by relying on more specific, concrete IA labels like ``Close-Minded Absolutism'' and ``Condescending Attitude'' to capture the high level idea of being `Closed to Diverse Perspectives'. \\
        \hline
        Embraces Mystery & --- & Accepts and appreciates the unknown or spiritual aspects beyond full comprehension. & --- & Political discourse, unlike religious discourse, tends to be rooted in empirical evidence, data, and concrete outcomes, rather than abstract concepts or ambiguities. Therefore, the idea of ``embracing mystery'' is more relevant to philosophical or spiritual contexts, where uncertainty and the unknown are accepted as part of deeper existential reflections, and not as applicable to political discourse. \\
        \hline
        Unsupported Claim & --- & Assertion that lacks evidence or adequate support, making it unreliable or unverifiable. & --- & In the context of politics, this label was deemed too subjective to serve as a consistent IA marker, especially because online political discourse is fast-moving in a way that makes sufficiently supporting all claims difficult. We viewed unsupported claims in the same vein as misinformation, which has a generally agreed upon definition but not a generally agreed upon method of identification \cite{altay_survey_2023}, and omitted the label to ensure our focus remained on intellectual humility and not fact-checking political debates. \\
        \hline
\end{tabular}
\hspace*{\fill}
\end{sidewaystable}
\clearpage
\twocolumn

\subsection{A.1: Human Annotators}
There are three human annotators for the process of refining the codebook and labeling the dataset. 

\subsubsection{Annotator \#1:} Undergraduate researcher studying computer science and philosophy. Native English speaker. 

\subsubsection{Annotator \#2:} Undergraduate researcher studying computer science and philosophy. Native English speaker. 

\subsubsection{Annotator \#3:} Graduate student studying interdisciplinary design and media. Native English speaker. 

\subsection{A.2: Annotator Agreement}
Table \ref{tab:kappa_table} shows the average Cohen's Kappa per sub-label, computed by averaging three different pairwise Cohen's Kappa values for each sub-label in each coding phase. 
\begin{table*}[ht] 
    \centering
    \begin{tabular}{lcccc} 
    \toprule
         \textbf{Label} & \textbf{Kappa (Phase 1)} & \textbf{Kappa (Phase 2)} & \textbf{Kappa (Phase 3)} \\
         \midrule
         Acknowledges Personal Beliefs & .37 & .67 & .87    \\
         Engages Respectfully with Diverse Perspectives & .26 & .33 & .38  \\
         Recognizes Limitations in ones own Knowledge or Beliefs & 0 & .33 & 1 \\
         Reconsiders Beliefs when Presented with New Evidence & 0 & 0 & 0  \\
         Seeks out New Information & 0 & .93 & .91  \\
         Displays Empathy & 0 & 0 & 0 \\
         Polarizing or Tribalistic Language & .25 & .41 & .67 \\
         Condescending Attitude & .27 & .14 & .36  \\
         Ad Hominem & .22 & 0 & .22 \\
         Displays Prejudice & 0 & 0 & 0\\
         Close-minded Absolutism & .20 & -.01 & .22 \\
         Overinflated Expertise & .33 & 0 & 0 \\
         \bottomrule
    \end{tabular}
    \caption{Summary of Cohen's Kappa scores throughout the iterative codebook refinement phase} 
    \label{tab:kappa_table}
\end{table*}

\section{Appendix B: IH Classifier Development}
\label{sec:classifierappendix}
\subsection{B.1: PerspectiveAPI}
\begin{figure}
    \centering
\includegraphics[width=.8\linewidth]{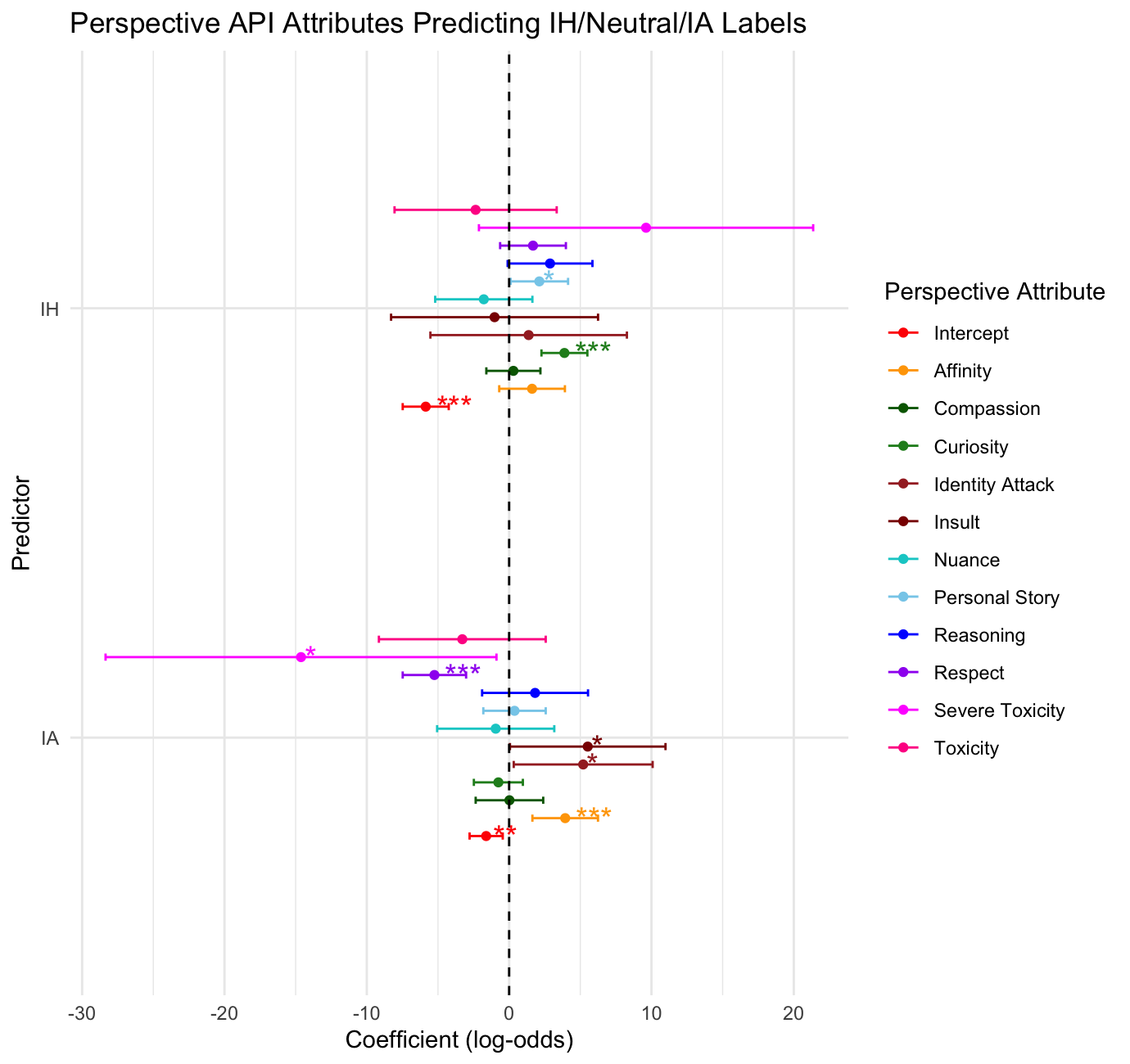}
    \caption{Results from the logistic regression attempting to understand which PerspectiveAPI labels were predictive of our IH/Neutral/IA Labels. *** indicates $p<.001$, ** indicates $p<.01$, * indicates $p<.05$.}
    \label{fig:perspective-logit}
\end{figure}

\begin{figure*}
    \centering
\includegraphics[width=1\linewidth]{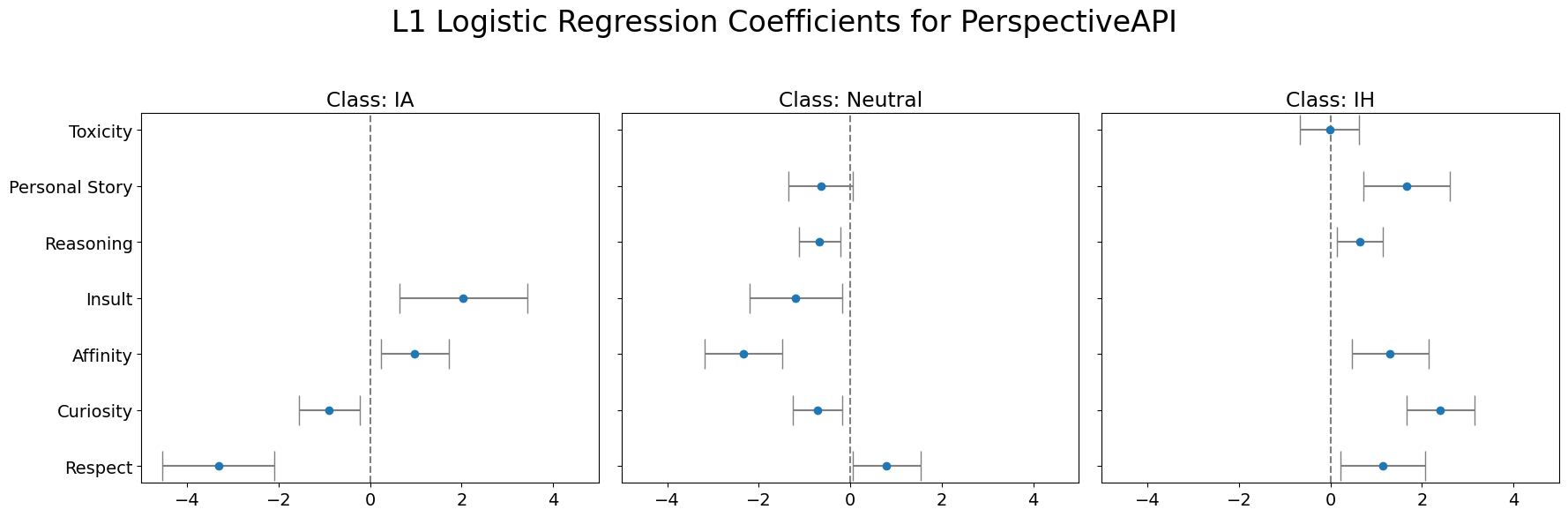}
    \caption{Non-Zero Coefficients from the L1 Logistic Regression for PerspectiveAPI feature selection. Error bars represent bootstrapped standard errors.}
    \label{fig:perspective-logl1}
\end{figure*}

Since PerspectiveAPI's attributes do not map neatly to our IH/IA definitions, we chose to use popular feature selection methods to identify significant attributes. First, we computed a logistic regression to understand which of the Perspective attributes (Affinity, Curiosity, Insult, Personal Story, Toxicity, and Respect) were predictive of our coarse IH/IA/Neutral labels. The results are shown for different models in Figure \ref{fig:perspective-logit}. Another popular method for feature selection is L1 (Lasso) Logistic regression. The results from this analysis are shown in Figure \ref{fig:perspective-logl1}.

Since the traditional logistic regression and L1 Logistic Regression yielded somewhat different results about which features were significant, we developed and tested different models to find the most effective one. We developed one model including only features that were statistically significant (at a significance level of $0.05$ in the traditional logistic regression (Affinity,  Curiosity, Identity Attack, Insult, Personal Story, Respect, Severe Toxicity). We also developed a model including only features that were non-zero in the L1 Logistic Regression (Affinity, Curiosity, Insult, Personal Story, Reasoning, Respect, Toxicity). We also developed a model including only features that were significant in both models (Affinity, Curiosity, Insult, Personal Story, Respect). The average F1 scores for all model are shown in Table \ref{tab:f1-results-perspective}. Since the results are comparable, we used the ``Significant in Both Regressions'' Model as our PerspectiveAPI operationalization. 

\begin{table*}[!]
    \centering
    \begin{tabular}{c|c|c|c|c}
         Model & IH & Neutral & IA & Weighted Average \\
         Significant in Logistic Regression & $.53 \pm .03$ & $.82 \pm .01$ & $.23 \pm .03$ & $.67 \pm .02$ \\
         Significant in L1 Regression & $.53 \pm .02$ & $.81 \pm .01$ & $.21 \pm .03$ & $.66 \pm .01$\\
         Significant in Both Regressions & $.53 \pm .03$ & $.82 \pm .01$ & $.23 \pm .04$ & $.67 \pm .02$
    \end{tabular}
    \caption{F1 Scores from different multinominal logistic regression models using PerspectiveAPI attributes. Significant in Logistic Regression included Compassion, Curiosity, Identity Attack, Insult, Personal Story, Respect, Severe Toxicity as features. Significant in L1 Regression included Affinity, Curiosity, Insult, Personal Story, Reasoning, Respect, Toxicity as features. Significant in Both Regressions included Affinity, Curiosity, Insult, Personal Story, Respect as features}
    \label{tab:f1-results-perspective}
\end{table*} 

\subsection{B.2: Large Language Model}
\subsection{Model Selection}
We tried using multiple models (GPT-4o, GPT-4, GPT-4.1, GPT-4.1-mini, GPT-4.1-nano, Claude Sonnet 4, and Claude Haiku 3.5) for this classification task. At the time, GPT 4.1 was the state of the art OpenAI model available and Claude Sonnet 4 / Claude Haiku 3.5 were the state of the art Anthropic models available. We noticed early on that reasoning models, such as GPT-4o, and smaller models, like GPT-4.1-nano, performed very poorly on this task and did not prioritize using them for subsequent testing. 
\subsubsection{Sublabels}
We tried a range of prompting techniques and a range of language models for the sub-label classification task. For prompting techniques, we started by just including the sub-labels and their associated definitions from the codebook. We then tried including the following components, both individually and in combination with one another, for each sub-label in the prompts:
\begin{itemize}
    \item Examples of posts from the dataset that had a specific sub-label applied
    \item Examples of posts from the dataset that did not have a specific sub-label applied
    \item The frequency of the sub-labels observed in our training set in an attempt to account for the imbalance of the labels in our dataset.
    \item The submission text the focal post was responding to and previous comments (when applicable) as additional context
    \item Shortening the names of the sublabels in an attempt to have the LLM rely more on our provided definitions than the actual names of the labels
\end{itemize}
We also attempted to provide additional contextual information by including the text the focal post was responding to and previous comments (when applicable), but found that this additional context degraded classifier performance.  

Additionally, we tried applying each sub-label individually using separate prompts. This means that for each focal post, we utilized 12 separate calls (with 12 separate prompts) to the AI model being tested. This did not improve accuracy.    

As stated, we tried these techniques across a number of AI models. For the sub-labels, across all prompting techniques and models, weighted F1 scores were $.52$ or less. After applying the sub-labels, we also determined a coarse label of IH/IA/Neutral based on the aggregation process described in the main text. We consistently had weighted F1 scores of $.65$ or less for this coarse labeling. Since these were in-line with results from PerspectiveAPI, we also chose to experiment with using language models to apply coarse labels directly. Specific examples of prompts and model combinations tried and resulting F1 scores is shown in Table \ref{tab:tested-prompts}.

\onecolumn
\begin{sidewaystable}[p]
\centering
\caption{A selection of some of the prompts and model combinations tested. Initially, we tried to mimic the human coding process by having the LLM classifier apply sublabels and then using the sublabels to apply IH/Neutral/IA labels. Prompts with that approach have both a Weighted F1 (IH, Neutral, IA) an Weighted F1 (Sublabels) score. Cost considerations related to using state of the art LLMs limited our ability to rigorously try every prompt/model pair and calculate error bars for every prompt/model pair tested. You'll see that for the final prompt attempt in the table, we were able to achieve a relatively high F1 score with GPT-4.1-mini. However, this higher F1 score was mainly driven by the 'Neutral' sublabel. In the final classifier, we chose to prioritize accuracy across all. Since prompt-engineering based LLM classifiers are relatively new, we tried to record all steps of our process for transparency to aid in future research. When applicable, examples are modified to protect privacy of Reddit users. When examples were provided, they were removed from the dataset for classifier validation. All prompts ended with the following standardized instructions: ``Respond with one of the following labels. Do not add any explanations. Valid labels: []. The text to label is: []". When additional context was provided, the standardized instructions included ``The context is: []".}
\label{tab:tested-prompts}
\vspace{0.5em}
\renewcommand{\arraystretch}{1.4}
\normalsize
\hspace*{\fill}
\begin{tabular}{>{\raggedright\arraybackslash}p{2.2cm}|>{\raggedright\arraybackslash}p{11cm}|>{\raggedright\arraybackslash}p{1.8cm}|>{\raggedright\arraybackslash}p{1.8cm}|>{\raggedright\arraybackslash}p{1.8cm}}
        Description & Prompt & Model & Weighted F1 (IH, Neutral, IA) & Weighted F1 (Sublabels)\\
        \hline
        Labeling sublabels.
        
        Definitions directly from codebook & You are a researcher studying intellectual humility.  You are assigned a coding task with the following codebook. None: No labels apply Acknowledges Personal Beliefs: Affirms individual convictions by speaking openly without contempt and/or uses first-person language to express an opinion or viewpoint without contempt. Engages Respectfully with Diverse Perspectives: Directly addresses and thoughtfully responds to differing perspectives in a way that acknowledges their validity or rationale. Recognizes limitations in one's own knowledge or beliefs: Acknowledges that one's political knowledge, beliefs, or information sources may be incomplete or subject to bias. Seeks out new information: Actively searches for new knowledge and perspectives on political issues or clarification on statements made and/or posing non-rhetorical questions. Polarizing or Tribalistic Language: Characterizes opposing groups as inherently evil, less human, or existential threats, creating an "us vs. them" narrative that undermines productive dialogue and fuels division. Condescending Attitude: Overbearing or dismissive behavior that undermines others' perspectives or intellect. Close-minded Absolutism: Using strong, definitive language to express convictions without engaging with or acknowledging diverse perspectives. & GPT-4.1-mini & .62 & .47 \\
        \hline
        \textit{Same as above} & \textit{Same as above} & GPT-4.1 & .69 & .50\\
        \hline
        \textit{Same as above} & \textit{Same as above} & Claude Haiku & .53 & .32\\
        \hline
        \textit{Same as above} & \textit{Same as above} & Claude Sonnet & .60 & .45\\
        \hline
        \textit{Same as above} & \textit{Same as above} & GPT-4o & .52 & .31\\
        \hline
\end{tabular}
\hspace*{\fill}
\end{sidewaystable}
\clearpage

\begin{sidewaystable}[p]
\centering
\renewcommand{\arraystretch}{1.4}
\normalsize
\hspace*{\fill}
\begin{tabular}{>{\raggedright\arraybackslash}p{2.2cm}|>{\raggedright\arraybackslash}p{11cm}|>{\raggedright\arraybackslash}p{1.8cm}|>{\raggedright\arraybackslash}p{1.8cm}|>{\raggedright\arraybackslash}p{1.8cm}}
        Description & Prompt & Model & Weighted F1 (IH, Neutral, IA) & Weighted F1 (Sublabels)\\
        \hline
        Sublabels. 
        
        Definitions slightly modified from codebook. 
        
        Additional guidelines about when to apply ``None'' & You are a researcher studying intellectual humility.  You are assigned a coding task with the following codebook. When in doubt, err on the side of applying None. None: No labels apply Acknowledges Personal Beliefs: Affirms individual convictions by speaking openly without contempt and/or uses first-person language to express an opinion or viewpoint without contempt. Engages Respectfully with Diverse Perspectives: Directly addresses and thoughtfully responds to differing perspectives in a way that acknowledges their validity or rationale. Recognizes limitations in one's own knowledge or beliefs: Explicitly acknowledges that one's political knowledge, beliefs, or information sources may be incomplete or subject to bias. Seeks out new information: Actively searches for new knowledge and perspectives on political issues or clarification on statements made and/or posing non-rhetorical questions. Polarizing or Tribalistic Language: Characterizes opposing groups as inherently evil, less human, or existential threats, creating an "us vs. them" narrative that undermines productive dialogue and fuels division. Mentioning a group in a negative way or mentioning the negative actions of a group are not enough to warrant this label. The comment must explicitly characterize the opposing group as evil, less human, or existential threats. Condescending Attitude: Overbearing or dismissive behavior that undermines others' perspectives or intellect. Look for overuse of sarcasm. Close-minded Absolutism: Using strong, definitive language to express convictions without engaging with or acknowledging diverse perspectives. & GPT-4.1-mini & .65 & .47 \\
        \hline
\end{tabular}
\hspace*{\fill}
\end{sidewaystable}
\clearpage

\begin{sidewaystable}[p]
\centering
\renewcommand{\arraystretch}{1.4}
\normalsize
\hspace*{\fill}
\begin{tabular}{>{\raggedright\arraybackslash}p{2.2cm}|>{\raggedright\arraybackslash}p{11cm}|>{\raggedright\arraybackslash}p{1.8cm}|>{\raggedright\arraybackslash}p{1.8cm}|>{\raggedright\arraybackslash}p{1.8cm}}
        Description & Prompt & Model & Weighted F1 (IH, Neutral, IA) & Weighted F1 (Sublabels)\\
        \hline
        Sublabels. 
        
        Definitions slightly modified. 
        
        Information about how frequently each label appears and examples for when to/when to not apply certain labels. & You are a researcher studying intellectual humility.  You are assigned a coding task with the following codebook. When in doubt, apply None. None: No labels apply. Acknowledges Personal Beliefs: Affirms individual convictions by speaking openly without contempt and uses first-person language to express an opinion or viewpoint. Do not apply this label if the comment is expressing an emotion. 16\ of posts have this label. This is an example of a post that is not Acknowledges Personal Beliefs: "He wasn't saying it for the right reasons, but Rubio was right: the newspapers are the enemy of the masses". Engages Respectfully with Diverse Perspectives: Directly addresses and thoughtfully responds to differing perspectives in a way that acknowledges their validity or rationale. 3\% of posts have this label. Examples of posts with this label are: "I'm not for it, but as a negotiating tactic it seems alright.", "I don't know. I think it can be necessary sometimes. I do agree that its importance is probably over stated." Recognizes limitations in one's own knowledge or beliefs: Explicitly acknowledges that one's political knowledge, beliefs, or information sources may be incomplete or subject to bias. 2\% of posts have this label. Seeks out new information: Actively searches for new knowledge and perspectives on political issues or clarification on statements made and/or posing non-rhetorical questions. 5\% of posts have this label. Polarizing or Tribalistic Language: Characterizes opposing groups as inherently evil, less human, or existential threats, creating an "us vs. them" narrative that undermines productive dialogue and fuels division. Mentioning a group in a negative way or mentioning the negative actions of a group are not enough to warrant this label. The comment must explicitly characterize the opposing group as evil, less human, or existential threats. 6\% of posts have this label. This is an example of a post that is not Polarizing or Tribalistic Language: "Clinton had two terms, that was two too many. He's a loser" Condescending Attitude: Overbearing or dismissive behavior that undermines others' perspectives or intellect directing explicitly towards the reader of the comment. Look for overuse of sarcasm. 7\% of posts have this label. Close-minded Absolutism: Using strong, definitive language to express convictions without engaging with or acknowledging diverse perspectives. 5\% of posts have this label. Examples of posts with this label are: "In my opinion, it is dumb and people make mountains out of molehills over nothing. You want to go celebrate a cultural holiday? Go have fun. Life is too short to get butthurt over small issues like this.", "It doesn't work, and anyone with basic math knowledge can see that for themselves.". This is an example of a post that is not Close-Minded Absolutism: "The only individuals who are on the fence about Haley are those waking up from a 15 year coma. All the others are lying." & GPT-4.1-mini & .62 & .49\\
        \hline
        \textit{Same as above} & \textit{Same as above} & GPT-4.1 & .71 & .52 \\
        \hline
\end{tabular}
\hspace*{\fill}
\end{sidewaystable}
\clearpage

\begin{sidewaystable}[p]
\centering
\renewcommand{\arraystretch}{1.4}
\normalsize
\hspace*{\fill}
\begin{tabular}{>{\raggedright\arraybackslash}p{2.2cm}|>{\raggedright\arraybackslash}p{11cm}|>{\raggedright\arraybackslash}p{1.8cm}|>{\raggedright\arraybackslash}p{1.8cm}|>{\raggedright\arraybackslash}p{1.8cm}}
        Description & Prompt & Model & Weighted F1 (IH, Neutral, IA) & Weighted F1 (Sublabels)\\
        \hline
        Shortened sublabels.

        Providing additional context (other posts in the thread) in addition to the focal post & You are a researcher studying intellectual humility. You have to do a coding task with the following dictionary. You will be given a post to code as well as the post that preceded it. The preceding post is only to supply additional context.None: None of the labels apply. If you apply this, do not apply any other labels. Acknowledges beliefs: Uses first-person language, like "I" or "My", to explicitly express an opinion or viewpoint without contempt.  Do not apply this label if someone is using first-person language to express an emotion, share an experience, or if the actual opinion is phrased in a passive or third-person voice. Engages respectfully:  Directly addresses and thoughtfully responds to differing perspectives in a way that acknowledges their validity or rationale.Recognizes limitations: Acknowledges that one's political knowledge, beliefs, or information sources may be incomplete or subject to bias.Seeks out new information: Actively searches for new knowledge and perspectives on political issues or clarification on statements made and/or posing non-rhetorical and not sarcastic questions.  The context may be particularly helpful here to understand sarcasm. Tribalistic: Characterizes opposing groups as inherently evil, less human, or existential threats, creating an "us vs. them" narrative that undermines productive dialogue and fuels division. Just mentioning a group negatively is not enough to apply this label, there must be an implication that the group mentioned is inherently evil or threatening. The label does not apply If the language is describing an action of the group, not the group themselves. Err on the side of not applying this label when in doubt.Condescending: Overbearing or dismissive behavior that undermines others' perspectives or intellect. To apply this label, the comment should be directed towards the reader of the comment, not a third-party absent from the conversation. Err on the side of not applying this label when in doubt. Close-minded: Using strong, definitive language to express convictions that broadly rejects other viewpoints or additional dialogue and projects their opinion as inevitable. Err on the side of not applying this label when in doubt.  & GPT-4.1-mini & .51 & .40\\
        \hline
\end{tabular}
\hspace*{\fill}
\end{sidewaystable}
\clearpage

\begin{sidewaystable}[p]
\centering
\renewcommand{\arraystretch}{1.4}
\normalsize
\hspace*{\fill}
\begin{tabular}{>{\raggedright\arraybackslash}p{2.2cm}|>{\raggedright\arraybackslash}p{11cm}|>{\raggedright\arraybackslash}p{1.8cm}|>{\raggedright\arraybackslash}p{1.8cm}|>{\raggedright\arraybackslash}p{1.8cm}}
        Description & Prompt & Model & Weighted F1 (IH, Neutral, IA) & Weighted F1 (Sublabels)\\
        \hline
        Coarse Labels.

        Inclusion of most labels that appeared in the codebook with modified definitions. & You are assigned a coding task with the following codebook. Label: IH Definition: Intellectual Humility Look for comments that contain signs of any of the following: Acknowledges Personal Beliefs: Affirms individual convictions by speaking openly without contempt and uses first-person language to express an opinion or viewpoint. Engages Respectfully with Diverse Perspectives: Directly addresses and thoughtfully responds to differing perspectives in a way that acknowledges their validity or rationale. Recognizes limitations in one's own knowledge or beliefs: Explicitly acknowledges that one's political knowledge, beliefs, or information sources may be incomplete or subject to bias. Seeks out new information: Actively searches for new knowledge and perspectives on political issues or clarification on statements made and/or posing non-rhetorical questions. Displays Empathy: Demonstrates an understanding of and sensitivity to other people in the argument's emotional experiences.Reconsiders beliefs when presented with new evidence: Demonstrates a willingness to rethink political beliefs when credible, new information challenges previous assumptions within the thread. Label: IA Definition: Intellectual Arrogance Look for comments that contain signs of any of the following: Polarizing or Tribalistic Language: Characterizes opposing groups as inherently evil, less human, or existential threats, creating an "us vs. them" narrative that undermines productive dialogue and fuels division. Mentioning a group in a negative way or mentioning the negative actions of a group are not enough to warrant this label. The comment must explicitly characterize the opposing group as evil, less human, or existential threats. Condescending Attitude: Overbearing or dismissive behavior that undermines others' perspectives or intellect directing explicitly towards the reader of the comment. Look for overuse of sarcasm. Close-minded Absolutism: Using strong, definitive language to express convictions without engaging with or acknowledging diverse perspectives. Ad Hominem: The argument attacks the person making the argument instead of addressing the argument itself. When applying this label, the comment must be directed towards the other person engaged in the conversation - it does not apply if someone says an ad hominem attack against a person not present in the conversation (for example a politician) Overinflated Expertise: Exaggerates one's own expertise or experience to make overly definitive or generalized assertions. Label: Neutral. Definition: Neither of the labels apply. When in doubt, apply this label. & GPT-4.1-mini & .66 & \textit{n/a} \\
        \hline
\end{tabular}
\hspace*{\fill}
\end{sidewaystable}
\clearpage
\twocolumn
\subsubsection{Coarse Labels}
As we experimented with different prompts for the fine-grained sub-label classifier, we determined that a coarse-grained classifier that simply determines whether or not a post should be classified as IH, IA, or Neutral would be sufficient for our intended downstream tasks. To arrive at a suitable prompt for this task, similar to the fine-grained task, we tried zero-shot prompting and few-shot prompting across a number of different language models. In general, given the simpler nature of the coarse-grained labeling task, performance across models tended to be higher. 

The final prompt used for the winning GPT-4.1 coarse classifier presented and used in the main text of the paper:
\begin{quote}
You are assigned a coding task with the following codebook.

Label: IH

Definition: Intellectual Humility
Look for comments that contain signs of any of the following:
Acknowledges Personal Beliefs: Affirms individual convictions by speaking openly without contempt and uses first-person language to express an opinion or viewpoint.
Engages Respectfully with Diverse Perspectives: Directly addresses and thoughtfully responds to differing perspectives in a way that acknowledges their validity or rationale.
Recognizes limitations in one’s own knowledge or beliefs: Explicitly acknowledges that one’s political knowledge, beliefs, or information sources may be incomplete or subject to bias.
Seeks out new information: Actively searches for new knowledge and perspectives on political issues or clarification on statements made and/or posing non-rhetorical questions.

Label: IA

Definition: Intellectual Arrogance
Look for comments that contain signs of any of the following:
Polarizing or Tribalistic Language: Characterizes opposing groups as inherently evil, less human, or existential threats, creating an "us vs. them" narrative that undermines productive dialogue and fuels division. Mentioning a group in a negative way or mentioning the negative actions of a group are not enough to warrant this label. The comment must explicitly characterize the opposing group as evil, less human, or existential threats.
Condescending Attitude: Overbearing or dismissive behavior that undermines others' perspectives or intellect directing explicitly towards the reader of the comment. Look for overuse of sarcasm.
Close-minded Absolutism: Using strong, definitive language to express convictions without engaging with or acknowledging diverse perspectives.

Label: Neutral

Neither of the labels apply. When in doubt, apply this label.

Respond with one of the following labels. Do not add any explanations.

Valid labels:
IH, IA, Neutral

The text to label is: \{Comment to Label\}
\end{quote}

\section{B.3: Error Analysis}
\begin{figure}
    \centering
    \includegraphics[width=1\linewidth]{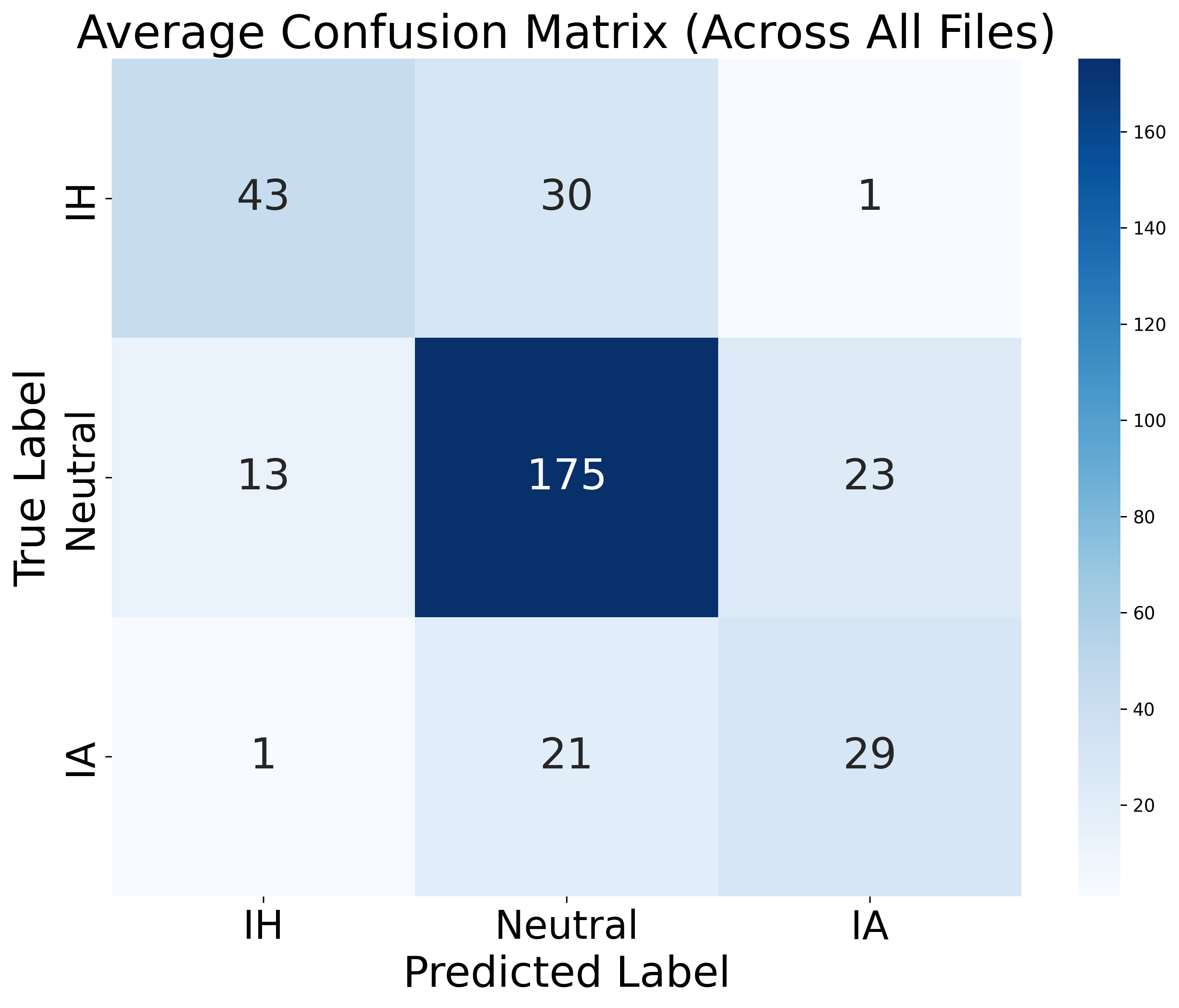}
    \caption{The average results from 20 trials with our GPT-4.1 best LLM classifier.}
    \label{fig:confusion-matrix}
\end{figure}

Table \ref{tab:classifier-error} shows a breakdown of the precision and recall for each class in our GPT-4.1 best LLM classifier. Figure \ref{fig:confusion-matrix} shows an average confusion matrix (across twenty labeling tries) for our GPT-4.1 best LLM classifier. The average Cohen's Kappa (across twenty labeling tries) between our classifier and our human labels is a .49, indicating moderate agreement. In our dataset of 359 posts, 23 posts were clearly made by bots (had the phrase *I am a bot  . . .*). These posts were removed from error analysis. Including these posts, as we did when calculating inter-human-rater agreement would only increase the Cohen's Kappa score to .50, still indicating moderate agreement.

Since we instructed our classifier to err on the side of ``Neutral", our classifier predictably has the most errors mislabeling IH and IA posts as neutral. However, since the errors are relatively symmetrically distributed (40\% of IA posts are labeled as Neutral and 40\% of IH posts are labeled as Neutral), there are very rarely misclassifications between IH/IA groups ($<$1.5\% of IH posts are labeled IA and $<$2\% of IA posts are labeled IH), and, as stated, there is a tendency to classify towards Neutral, we believe the directional findings of both our Reddit Analysis and RCT Analysis are valid and the effect magnitudes in both may be underestimates. The impact of measurement error may be further mitigated by the fact that all dependent variables in our analysis are averages of user posts, meaning the symmetric measurement error may cancel out (although the impact of this depends on the number of each type of post made per user). 

With that being said, we still looked through some specific misclassifications to qualitatively understand potential improvements. There are a significant number of ``Neutral" posts that were labeled ``IA". Looking through informally, some of them contain quotations of other posts (which may or may not be ``IA"). 
For example (post modified to protect user privacy):

``$>$This punk is so sad it defies logic. I cannot bear to see or hear this asshole. What did we do to deserve this punishment?

IKR"

Human annotators likely noticed that $>$ indicates a reply to an existing post. Although that quoted post displays IA tendencies, the response by the author of the focal post is not IA. This issue is not likely to be relevant for the randomized control trial, but is relevant for the Reddit analysis. Future research should better handle instances of quoted posts in focal posts.

Some of these posts were labeled with IA labels that were dropped from analysis, such as the example below: ``OP is a Pedo supporter. Disgusting man. Get a life.", which had the ``Ad hominem" label in the original coding but, since that label was dropped from final consideration due to low frequency, ended up labeled as ``Neutral" in the human dataset. This indicates poor handling of dropped sublabels by the human annotators. In our released dataset, all sublabels will be applied and future researchers can determine how they wish to handle such scenarios.
\begin{table}
\centering
\begin{tabular}{lccc}
\toprule
Class & Precision & Recall & F1-Score \\
\midrule

IH & 0.76 $\pm$ 0.02 & 0.59 $\pm$ 0.02 & 0.66 $\pm$ 0.02 \\
Neutral & 0.78 $\pm$ 0.01 & 0.83 $\pm$ 0.01 & 0.80 $\pm$ 0.01 \\
IA & 0.55 $\pm$ 0.01 & 0.58 $\pm$ 0.02 & 0.56 $\pm$ 0.01 \\
\bottomrule
\end{tabular}
\caption{Per-Class Performance Metrics (Mean $\pm$ Std)}
\label{tab:classifier-error}
\end{table}

\section{Appendix C: Reddit Analysis}
\label{sec:redddit-analysis}
\subsection{C.1 Environment Classification}
Table \ref{tab:subreddit_environment} presents the ten subreddits analyzed and their subreddit-level mean IH scores computed on the March–April sample used for environment classification. We assign labels by the sign of the mean: subreddits with positive means are classified as IH environments, and those with negative means as IA environments.
\begin{table}[h]
    \centering
    \renewcommand{\arraystretch}{1.2}
    \begin{tabular}{l c c}
    \hline
    Subreddit & Mean IH score & Environment \\
    \hline
    Anarchism           &  0.188  & IH environment \\
    moderatepolitics    &  0.137  & IH environment \\
    socialism           &  0.112  & IH environment \\
    PoliticalDiscussion &  0.012  & IH environment \\
    Libertarian         & -0.001  & IA environment \\
    Foodforthought      & -0.128  & IA environment \\
    Conservative        & -0.129  & IA environment \\
    politics            & -0.130  & IA environment \\
    democrats           & -0.136  & IA environment \\
    Republican          & -0.172  & IA environment \\
    \hline
    \end{tabular}
    \caption{Mean IH scores by subreddit and their environment classification.}
    \label{tab:subreddit_environment}
\end{table}

To select these ten subreddits, we first conducted a manual review to identify eligible subreddits related to U.S. political discussion. This included location-based subreddits, like r/AmericanPolitics, and ideology-based subreddits that referenced American politics, like r/socialism. In this case, 'reference' meant the subreddit had to mention the U.S. in their descriptions or have a post explicitly about US politics from the last month (at the time of the search). Photo-based subreddits, such as r/PoliticalHumor, private subreddits, and subreddits that did not have a post within the last year were excluded from consideration. After compiling a list of 97 eligible subreddits, we selected the top ten based on number of active users. While we had access to Reddit usernames and other information posted publicly on Reddit, in an effort to preserve privacy, we did not conduct in-depth investigations into the behaviors of any specific users. 

\subsection{C.2: Details for Cross-Environment Users}
Table~\ref{tab:user_group} shows additional details about the cross-environment users, including user count and comment count for each user group.

\begin{table}[H]
    \centering
    \begin{tabular}{ccc}
        \toprule
         \textbf{User Group} & \textbf{User Count} & \textbf{Comment Count} \\
         \midrule
         Top 10\% & 189 & 19,684  \\
         Top 25\% & 644 & 35,842  \\
         Top 50\% & 1,435 & 55,278  \\
         Top 100\% (All) & 3,640 & 68,874 \\
         \bottomrule
    \end{tabular}
    \caption{Cross-environment users by activity level.}
    \label{tab:user_group}
\end{table}

We compare each cross-environment user’s average IH score across the two environments using a paired \textit{t}-tests and, as a non-parametric robustness check, Wilcoxon signed-rank tests. The null hypothesis is that there is no difference in a user’s mean IH score between IH and IA environments; the alternative is that the means differ. Table \ref{tab:paired_ttest_results} reports results by activity subset. Positive mean difference values indicate that, on average, the same users’ comments score higher in IH subreddits than in IA subreddits, suggesting that users tend to express more intellectually humble content in IH environments.
\begin{table}[h]
    \centering
    \scriptsize
    \renewcommand{\arraystretch}{1.2}
    \begin{tabular}{p{1cm} p{0.5cm} p{0.7cm} p{0.3cm} p{0.5cm} p{0.3cm} p{0.7cm} p{0.3cm} p{0.5cm}}
    \hline
        User group & User count & Comment count & Mean diff & t-statistic & $p$-value (t) & Wilcoxon statistic & $p$-value (w) & Cohen's $d$ \\
        \hline
        Top 10\%  &   189 & 19,684 & 0.08 & 5.18 & *** & 5,508    & *** & 0.38 \\
        Top 25\%  &   644 & 35,842 & 0.09 & 8.88 & *** & 59,565   & *** & 0.35 \\
        Top 50\%  & 1,435 & 55,278 & 0.10 & 10.69 & *** & 297,318  & *** & 0.28 \\
        Top 100\% & 3,640 & 68,874 & 0.08 & 9.41 & *** & 1,577,606 & *** & 0.16 \\
    \hline
    \end{tabular}
    \caption{Paired t-test and Wilcoxon signed-rank test results across user groups. 
    $^{***}p<0.001$, $^{**}p<0.01$, $^{*}p<0.05$.}
    \label{tab:paired_ttest_results}
\end{table}

\subsection{C.3: Ordinal Logistic Regression: Specification and Full Results}
\label{sec:OLS-results}
We fit five ordinal logistic specifications incrementally; the specifications are summarized in Table \ref{tab:Ordered Logit models}. 

\begin{table}[H]
    \centering
    \begin{tabular}{p{1.2cm}|p{7cm}}
        & Formulation \\
        \hline
        Model 1 & ${IH}_{i} \sim {Env}$ \\
        Model 2 & ${IH}_{i} \sim {Env} + {Author}$ \\
        Model 3 & ${IH}_{i} \sim {Env} + {Author} + {Topic}$ \\
        Model 4 & ${IH}_{i} \sim {Env} + {Author} + {Topic} + {RollingMean}_{3}$ \\
        Model 5 & ${IH}_{i} \sim {Env} + {Author} + {Topic} + {RollingMean}_{5}$ \\
    \end{tabular}
    \caption{Model specifications for ordered logistic regression.}
    \label{tab:Ordered Logit models}
\end{table}
Model 1 includes only the subreddit environment indicator; Model 2 adds author fixed effects; Model 3 adds topic fixed effects; Models 4 and 5 additionally include a thread-level rolling mean of prior IH scores based on the previous three and five comments, respectively. This sequence is used to examine how the estimated environment effect changes as controls are introduced and to compare subreddit-level ("macro") with thread-level ("micro") context.

Table~\ref{tab:olr_full_results} reports the complete estimates for Models~1–5 across activity subsets. As summarized in Section \ref{sec:redddit-Analysis and results}, the environment coefficient is positive and statistically significant in every specification; its decrease as author, topic, and thread-level controls are introduced. We also provide odds ratios for interpretability alongside coefficients and confidence intervals.

Across all user subsets, there appeared to be a stronger association between the 5 most recent preceding comments and the IH level demonstrated in the focal comment, versus the 3 most recent preceding comments and the focal comment's demonstrated IH---suggesting that a slightly broader conversational context exerts a stronger influence on user behavior. For example, in the Top 50\% user group, the rolling mean based on the three previous comments had a coefficient of $\beta \approx 0.20$ (OR $\approx 1.22$), whereas the rolling mean based on the five most recent comments rose to $\beta \approx 0.28$, corresponding to OR $\approx 1.32$. Additionally, these associations appeared to strengthen for more active users.

\begin{table*}[ht!]
    \centering
    \renewcommand{\arraystretch}{1.2}
    \resizebox{\textwidth}{!}{%
    \begin{tabular}{l l r r r r r r r r r r r r r}
        \hline
        User Group & Model & N & Coef (Env) & p (Env) & OR (Env) & p (Env, Bonf.) & CI Low (Env) & CI Up (Env) & Coef (Roll) & p (Roll) & OR (Roll) & CI Low (Roll) & CI Up (Roll) & Pseudo $R^2$ \\
        \hline
        Top 10\%  & Model 1 & 12440 & 0.5040 & *** & 1.6554 & *** & 0.4279 & 0.5802 & --     & --  & --     & --     & --     & 0.0081 \\
                  & Model 2 & 12440 & 0.3309 & *** & 1.3922 & *** & 0.2329 & 0.4289 & --     & --  & --     & --     & --     & 0.0710 \\
                  & Model 3 & 12440 & 0.4377 & *** & 1.5492 & *** & 0.3093 & 0.5662 & --     & --  & --     & --     & --     & 0.0920 \\
                  & Model 4 & 11097 & 0.4027 & *** & 1.4958 & *** & 0.2648 & 0.5405 & 0.1748 & **  & 1.1910 & 0.0571 & 0.2926 & 0.0905 \\
                  & Model 5 & 11097 & 0.3893 & *** & 1.4760 & *** & 0.2510 & 0.5276 & 0.2616 & *** & 1.2990 & 0.1218 & 0.4014 & 0.0907 \\
        \hline
        Top 25\%  & Model 1 & 22944 & 0.5920 & *** & 1.8077 & *** & 0.5356 & 0.6485 & --     & --  & --     & --     & --     & 0.0110 \\
                  & Model 2 & 22944 & 0.3894 & *** & 1.4761 & *** & 0.3142 & 0.4646 & --     & --  & --     & --     & --     & 0.0886 \\
                  & Model 3 & 22944 & 0.4135 & *** & 1.5122 & *** & 0.3190 & 0.5081 & --     & --  & --     & --     & --     & 0.1020 \\
                  & Model 4 & 20791 & 0.3743 & *** & 1.4540 & *** & 0.2735 & 0.4751 & 0.2142 & *** & 1.2388 & 0.1274 & 0.3009 & 0.1039 \\
                  & Model 5 & 20791 & 0.3632 & *** & 1.4379 & *** & 0.2621 & 0.4643 & 0.2987 & *** & 1.3481 & 0.1956 & 0.4018 & 0.1041 \\
        \hline
        Top 50\%  & Model 1 & 35001 & 0.5629 & *** & 1.7558 & *** & 0.5165 & 0.6094 & --     & --  & --     & --     & --     & 0.0097 \\
                  & Model 2 & 35001 & 0.4237 & *** & 1.5276 & *** & 0.3584 & 0.4890 & --     & --  & --     & --     & --     & 0.0926 \\
                  & Model 3 & 35001 & 0.4285 & *** & 1.5350 & *** & 0.3486 & 0.5084 & --     & --  & --     & --     & --     & 0.1042 \\
                  & Model 4 & 31627 & 0.3899 & *** & 1.4768 & *** & 0.3049 & 0.4749 & 0.2025 & *** & 1.2244 & 0.1320 & 0.2730 & 0.1078 \\
                  & Model 5 & 31627 & 0.3789 & *** & 1.4607 & *** & 0.2937 & 0.4641 & 0.2827 & *** & 1.3267 & 0.1989 & 0.3665 & 0.1081 \\
        \hline
        Top 100\% & Model 1 & 43709 & 0.5486 & *** & 1.7308 & *** & 0.5070 & 0.5902 & --     & --  & --     & --     & --     & 0.0093 \\
                  & Model 2 & 43709 & 0.4168 & *** & 1.5172 & *** & 0.3601 & 0.4736 & --     & --  & --     & --     & --     & 0.0853 \\
                  & Model 3 & 43709 & 0.4118 & *** & 1.5096 & *** & 0.3420 & 0.4817 & --     & --  & --     & --     & --     & 0.0962 \\
                  & Model 4 & 39656 & 0.3672 & *** & 1.4437 & *** & 0.2930 & 0.4414 & 0.1920 & *** & 1.2117 & 0.1290 & 0.2549 & 0.0994 \\
                  & Model 5 & 39656 & 0.3582 & *** & 1.4308 & *** & 0.2838 & 0.4327 & 0.2548 & *** & 1.2903 & 0.1802 & 0.3295 & 0.0995 \\
        \hline
    \end{tabular}
    }
    \caption{Ordered logit regression results across user groups. 
    $^{***}p<0.001$, $^{**}p<0.01$, $^{*}p<0.05$. OR = odds ratio; CI = 95\% confidence interval. 
    Cells with `--` indicate that the variable was not included in the model. 
    The column ``p (Env)'' reports significance based on unadjusted $p$-values, 
    while ``p (Env, Bonf.)'' reports significance based on Bonferroni-adjusted $p$-values for the IH environment coefficient.}
    \label{tab:olr_full_results}
\end{table*}

\section{Appendix D: Intervention Specifications}
\begin{figure*}[t]
    \centering
    \includegraphics[width=\linewidth]{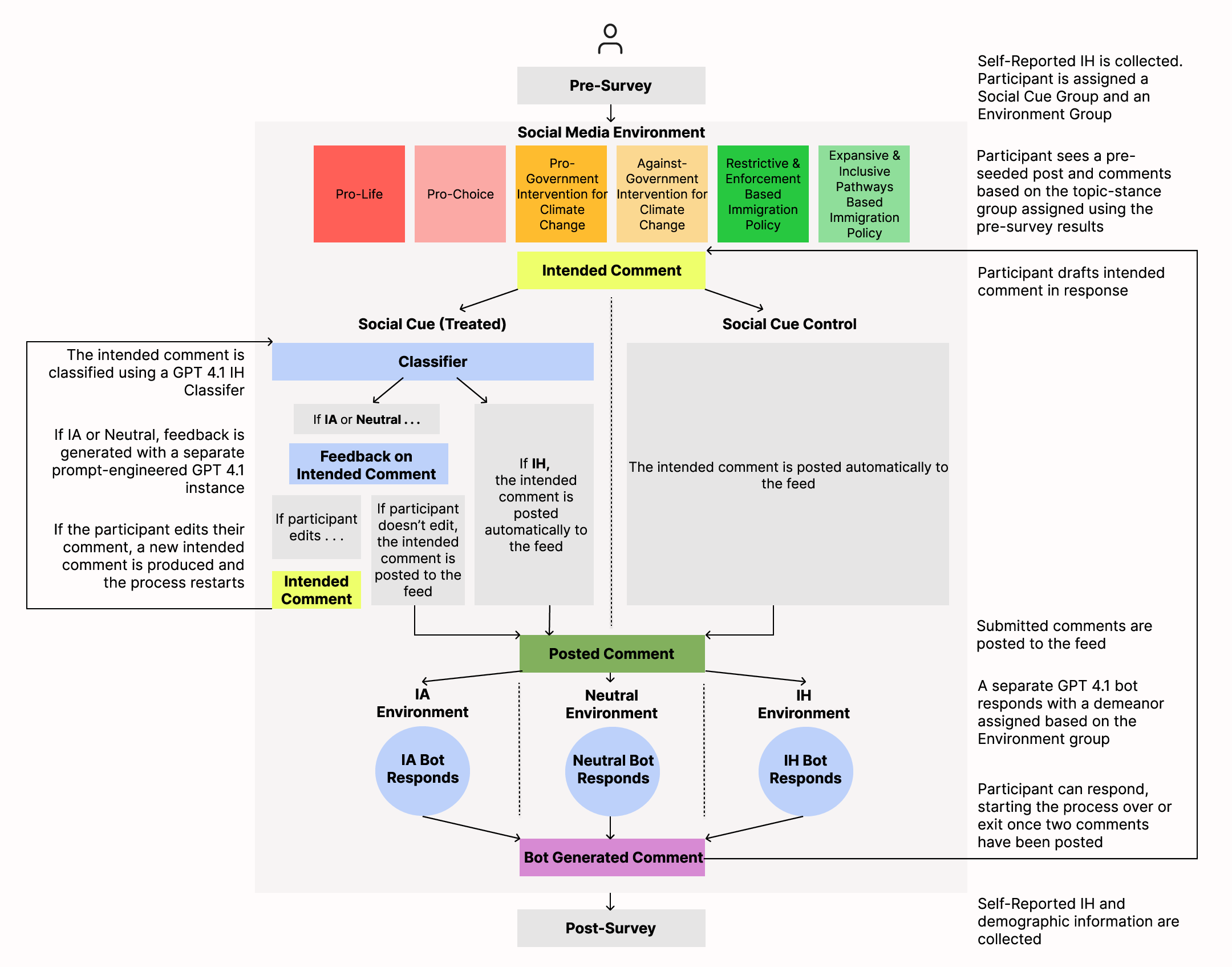}
    \caption{This diagram represents the lab-based experiment design flow. Randomization happens when participants enroll in the study. Participants stay in the same ``Social Cue" and ``Environment" group throughout the duration of the experiment.}
    \label{fig:intervention-flow}
\end{figure*}

Figure~\ref{fig:intervention-flow} demonstrates the experiment flow.

\label{sec:internvention}
\subsection{D.1: Self-Reported IH Survey Questions}
The following IH questions were lifted from the Comprehensive Intellectual Humility Scale \cite{krumrei-mancuso_development_2016} and used to calculate self-reported intellectual humility. Participants rated each question on a 10-point Likert scale with 1 being strongly disagree and 10 being strongly agree. Reverse-coded items are indicated with an asterisk.
\begin{itemize}
       \item When I am really confident in a belief, there is very little chance that belief is wrong.*
    \item I'd rather rely on my own knowledge about most topics than turn to others for expertise.*
    \item I am open to revising my important beliefs in the face of new information.
    \item I welcome different ways of thinking about important topics.
    \item I am willing to hear others out, even if I disagree with them in important ways.
    \item I can respect others, even if I disagree with them in important ways.
    \item I can have great respect for someone, even when we don't see eye-to-eye on important topics.
    \item Even when I disagree with others, I can recognize that they have sound points.
\end{itemize}

\subsection{D.2: Informed Consent for Intervention}
Each participant was shown the following text in the beginning of the experiment and had the opportunity to stop participating at any time. 
\begin{quote}
    Welcome to the Research Study!

You must be 18 years or older to participate in this study.

        This project is part of a [Redacted for Blind Review] research study
        exploring patterns of online discourse. By accepting this task, you will
        be enrolled in this study. Your answers to survey questions and other
        information that is produced as a result of your participation will
        remain anonymous and only available to members of the research team. As
        a part of this research study, you will be asked to interact with posts
        and users in a simulated social media environment to help illuminate how
        the design of online conversations can impact discussion quality. Please
        avoid submitting any identifiable information while using the simulated
        social media environment. The entire experience should take you around
        10 minutes.

        You will be compensated as specified in Prolific for participating in
        this study. To protect the integrity of the research, you will not be
        made aware of the precise reasons for being exposed to such information
        and experiences until after the survey. You can reach out to the Primary
        Investigator of this research study after you complete your
        participation to ask any questions or seek additional insights into the
        study itself.

        The research team may analyze information produced through this research
        to prepare and submit research papers to computational social science
        venues. Your de-identified information could be used for future research
        without additional informed consent. You will be asked to engage in
        political discussions, which may include topics such as climate change,
        reproductive rights, and immigration

        Your participation is entirely voluntary, and you can stop at any time.
        Please reach out to the Primary Investigator, Nabeel Gillani, with any
        questions or comments: n.gillani@northeastern.edu.

        If you have questions or concerns about your rights as a participant,
        please contact Northeastern University’s Human Research Protection
        Program at IRBReview@northeastern.edu.
\end{quote}
\subsection{D.3: Social Media Testing Environment}
\begin{figure}
    \centering
    \includegraphics[width=1\linewidth]{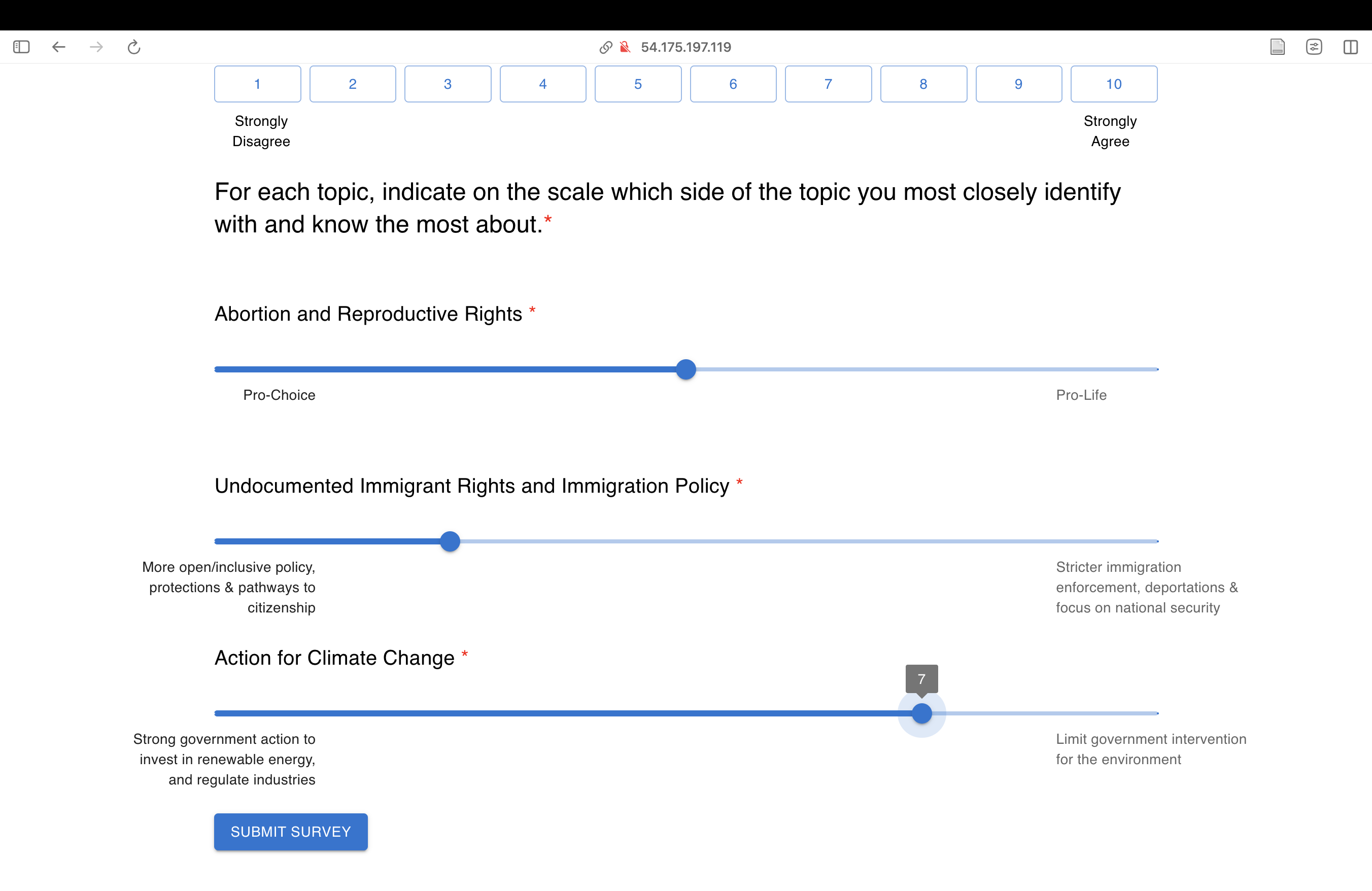}
    \caption{Questions used to gauge participants topic/stance on three issues. }
    \label{fig:pre-surveytopic}
\end{figure}

\begin{figure}
    \centering
    \includegraphics[width=1\linewidth]{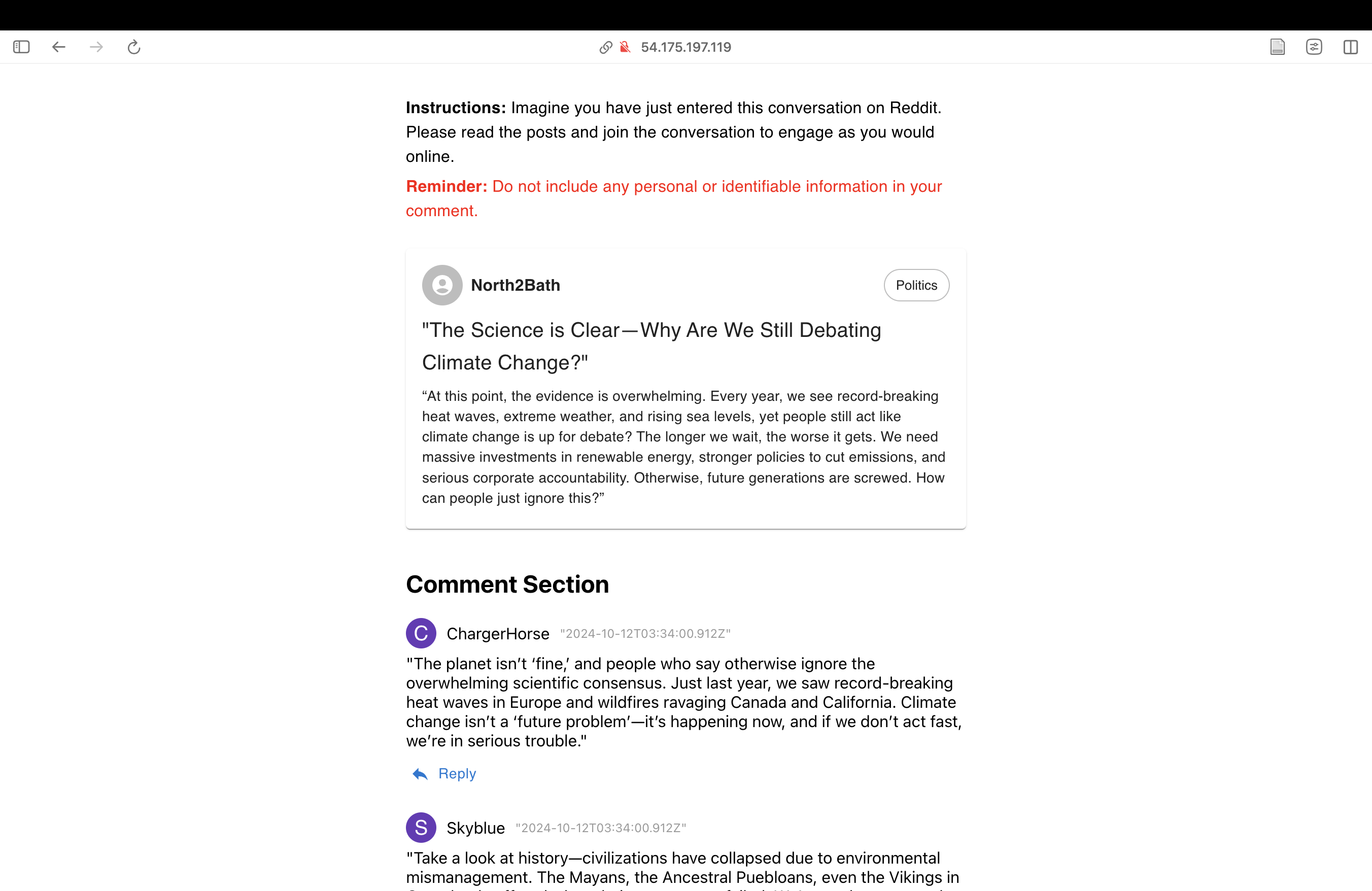}
    \caption{Example Reddit-like conversation interface participants saw in the test environment.}
    \label{fig:reddit-environment}
\end{figure}

\begin{figure}
    \centering
    \includegraphics[width=1\linewidth]{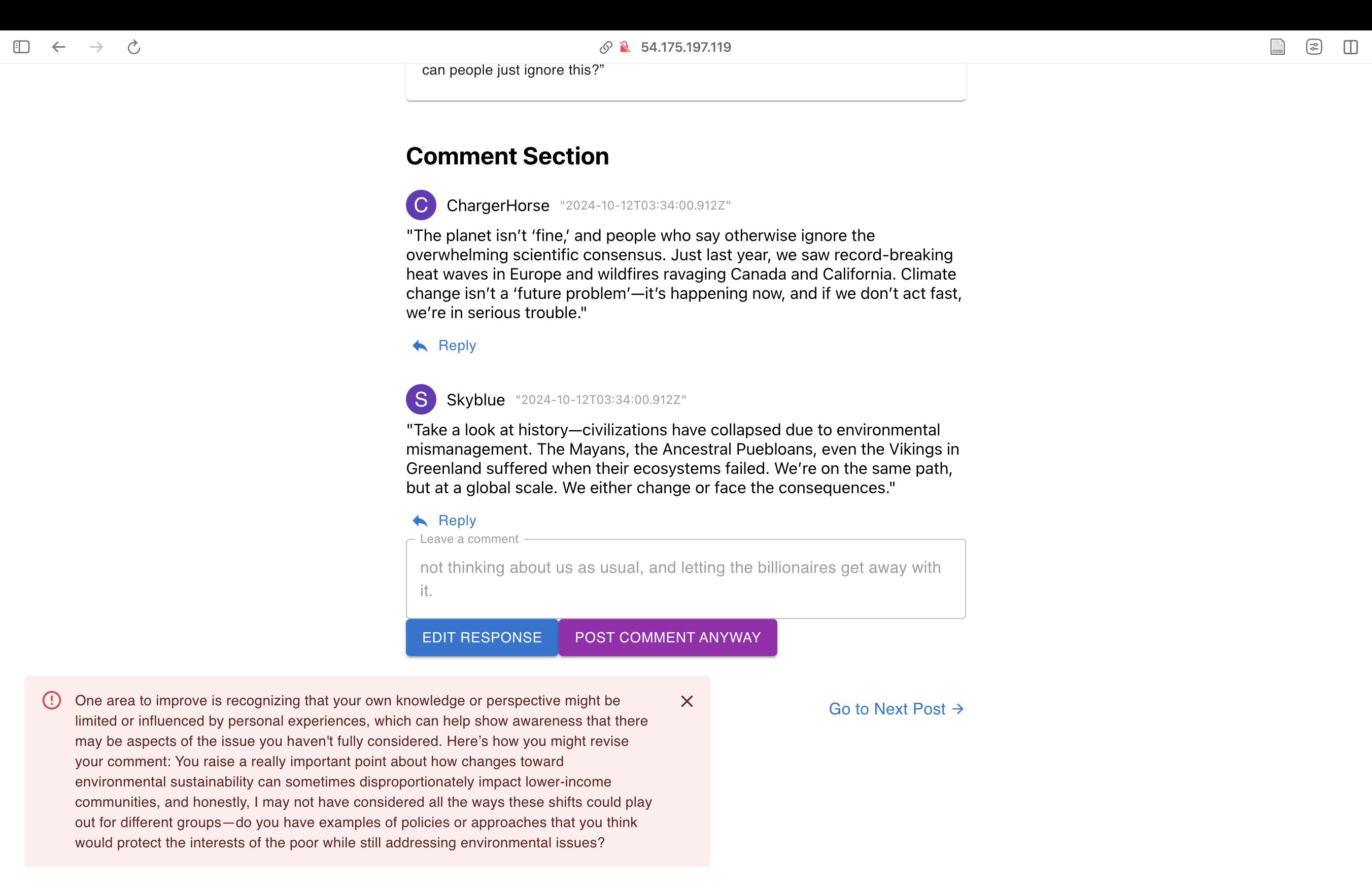}
    \caption{An example of the comment functionality and ``Social Cue Nudge'' intervention.}
    \label{fig:intervention-ui}
\end{figure}

To determine topic/stance group, participants answered the questions seen in Figure \ref{fig:pre-surveytopic}. Participants were assigned to the topic they expressed the most extreme opinion towards and assigned to conversations of opposite stance than they expressed. In the case of a tie, random selections were made. The Reddit-like conversation interface is shown in Figure \ref{fig:reddit-environment} and example of the comment functionality and intervention is shown in Figure \ref{fig:intervention-ui}.
\subsection{D.4: Personal Story Profiles of AI Dialogue Agents}
The following were the persona profiles of the AI bots participants interacted with. A random persona was selected to respond to participants' comments. In addition to the personas, information about the task was included in the prompt to ChatGPT. 
\begin{itemize}
    \item BlueSkyRider is a 45 year old white lady who grew up in California all her life in the same small house. Often supportive and curious, with ties to a non-profits helping people live more sustainably. She has 2 sons in sports.
    \item Grove is a 52-year-old African American woman who was raised in New York City. Growing up in a diverse environment, she values community resilience, economic opportunity, and accessible healthcare for all. She likes rock, references pop-culture often. She has a bulldog.
    \item Rambler is a 24 year old man who grew up with two moms in Wisconsin, Madison but hates the cold. He was in public school all his life before going to a private university. He enjoys cooking his own food and cares about grocery prices.
    \item Cake12 is a 29-year-old Latino man who grew up in Chicago. He works in high finance. He has a wife who is outspoken in mental health awareness and is a teacher in the suburbs of Chicago.
    \item ForestWish is a 37-year-old Korean-American woman who was born in Seoul but grew up in Texas. She is a software engineer who loves hiking, experimenting with fusion recipes, and is passionate about increasing diversity in STEM fields. She has a rescue cat named Mochi.
	\item iPlum is a 63-year-old retired Native American man from Arizona. He spent his career as a park ranger and advocates for land conservation and indigenous rights. He enjoys woodworking, storytelling, and teaching his grandchildren traditional crafts.
	\item FishPause is a 50-year-old Middle Eastern immigrant who came to the U.S. as a teenager. He owns a small café and values community-building through food. He is deeply involved in mentoring young entrepreneurs and spends weekends watching soccer with his extended family.",
	\item Bough23 is a 22-year-old non-binary person from Portland, Oregon. They are an artist who creates murals focusing on social justice themes and queer visibility. They ride their bike everywhere, work at a local bookstore, and are a huge fan of indie music festivals.
	\item HudsonLake0 is a 41-year-old Black man who grew up in Atlanta and served in the military. Now a firefighter, he takes pride in helping his community and mentoring young recruits. He enjoys classic jazz, barbecuing on weekends, and fixing up old motorcycles.
	\item KmThicket is a 33-year-old Indian-American woman from New Jersey. She is a pediatric nurse who believes in making healthcare more accessible for underserved communities. She practices yoga, enjoys Bollywood movies, and hosts dinner parties where she shares her grandmother’s family recipes.
\end{itemize}
\subsection{D.5: Intervention Prompts}
The following prompt was used for the ``Social Cue'' intervention. 
\begin{quote}
You are trying to help someone respond to a comment on Reddit with more intellectual humility. Intellectual humility consists of the following characteristics:
Acknowledges Personal Beliefs: Affirms individual convictions by speaking openly without contempt and/or uses first-person language to express an opinion or viewpoint without contempt.
Engages Respectfully with Diverse Perspectives: Directly addresses and thoughtfully responds to differing perspectives in a way that acknowledges their validity or rationale.
Recognizes limitations in one’s own knowledge or beliefs: Explicitly acknowledges that one’s political knowledge, beliefs, or information sources may be incomplete or subject to bias.
Seeks out new information: Actively searches for new knowledge and perspectives on political issues or clarification on statements made and/or posing non-rhetorical questions.
Pick one of these characteristics that could be improved in the message.
In one sentence, suggest the characteristic that they can improve. Don't just say the characteristic name and instead, explain what it is.
Suggest how their post can be improved by rewriting their post so that it incorporates the characteristic. Address the user directly.
Simply restate what they user is trying to say. Don't include any extra information.
Here is the message the person is the post the user is responding to: \{Participant Comment\}
\end{quote}
\subsection{D.6: Environment Experiment}
The demeanor prompt for bots in the Intellectually Humble Environment was:
\begin{quote}
    Act in an intellectually humble way by incorporating any of the following attributes into your response: Acknowledges Personal Beliefs: Affirms individual convictions by speaking openly without contempt and uses first-person language to express an opinion or viewpoint. Engages Respectfully with Diverse Perspectives: Directly addresses and thoughtfully responds to differing perspectives in a way that acknowledges their validity or rationale. Recognizes limitations in one’s own knowledge or beliefs: Explicitly acknowledges that one’s political knowledge, beliefs, or information sources may be incomplete or subject to bias. Seeks out new information: Actively searches for new knowledge and perspectives on political issues or clarification on statements made and/or posing non-rhetorical questions.
\end{quote}

The demeanor prompt for bots in the Intellectually Arrogant Environment was:
\begin{quote}
    Act in an intellectually arrogant way by incorporating any of the following attributes into your response: Polarizing or Tribalistic Language: Characterizes political opponents as inherently evil, less human, or existential threats, creating an "us vs. them" narrative that undermines productive dialogue and fuels division. Condescending Attitude: Overbearing or dismissive behavior that undermines others' perspectives or intellect. Close-minded Absolutism: Using strong, definitive language to express convictions without engaging with or acknowledging diverse perspectives.
\end{quote}

The demeanor prompt for bots in the Neutral Environment was:
\begin{quote}
   Act in neutral way. Do not include any of the following in your responses: Polarizing or Tribalistic Language: Characterizes political opponents as inherently evil, less human, or existential threats, creating an "us vs. them" narrative that undermines productive dialogue and fuels division. Condescending Attitude: Overbearing or dismissive behavior that undermines others' perspectives or intellect. Close-minded Absolutism: Using strong, definitive language to express convictions without engaging with or acknowledging diverse perspectives. Acknowledges Personal Beliefs: Affirms individual convictions by speaking openly without contempt and uses first-person language to express an opinion or viewpoint. Engages Respectfully with Diverse Perspectives: Directly addresses and thoughtfully responds to differing perspectives in a way that acknowledges their validity or rationale. Recognizes limitations in one’s own knowledge or beliefs: Explicitly acknowledges that one’s political knowledge, beliefs, or information sources may be incomplete or subject to bias. Seeks out new information: Actively searches for new knowledge and perspectives on political issues or clarification on statements made and/or posing non-rhetorical questions. 
\end{quote}
\section{Appendix E: Intervention Results}
\label{sec:appendix-results}
\subsection{E.1: Full Model Results}
The full  model results for the ``Base'' Models are shown in Table \ref{tab:full-base-model} and the full model results for the``Interaction'' Models are shown in Table \ref{tab:full-interaction-model}.

\begin{table*}
\begin{center}
\begin{tabular}{l c c c}
\hline
 & Demonstrated IH & Change in Self Reported IH & Number of Comments \\
\hline
(Intercept)             & $0.23^{***}$ & $-0.09$      & $6.99^{***}$ \\
                        & $(0.04)$     & $(0.07)$     & $(0.31)$     \\
Social Cue Nudge  & $0.25^{***}$ & $0.00$       & $-0.13$      \\
                        & $(0.04)$     & $(0.07)$     & $(0.31)$     \\
IA Environment & $-0.12^{*}$  & $-0.23^{**}$ & $0.35$       \\
                        & $(0.05)$     & $(0.09)$     & $(0.38)$     \\
IH Environment & $0.10^{*}$   & $0.08$       & $0.71$       \\
                        & $(0.05)$     & $(0.09)$     & $(0.39)$     \\
\hline
R$^2$                   & $0.13$       & $0.04$       & $0.01$       \\
Adj. R$^2$              & $0.13$       & $0.03$       & $0.00$       \\
Num. obs.               & $355$        & $355$        & $355$        \\
\hline
\multicolumn{4}{l}{\scriptsize{$^{***}p<0.001$; $^{**}p<0.01$; $^{*}p<0.05$}}
\end{tabular}
\end{center}
\caption{Full regression results for the ``Base'' Models presented for the three outcomes of interest.}
\label{tab:full-base-model} 
\end{table*}

\begin{table*}
\begin{center}
\begin{tabular}{l c c c}
\hline
 & Demonstrated IH & Change in Self Reported IH & Number of Comments \\
\hline
(Intercept)                                    & $0.28^{***}$ & $-0.15$  & $7.19^{***}$ \\
                                               & $(0.05)$     & $(0.09)$ & $(0.37)$     \\
Social Cue Nudge                         & $0.15^{*}$   & $0.13$   & $-0.54$      \\
                                               & $(0.07)$     & $(0.12)$ & $(0.53)$     \\
IA Environment                        & $-0.19^{**}$ & $-0.10$  & $0.02$       \\
                                               & $(0.07)$     & $(0.12)$ & $(0.52)$     \\
IH Environment                        & $0.02$       & $0.13$   & $0.41$       \\
                                               & $(0.08)$     & $(0.13)$ & $(0.55)$     \\
Social Cue Nudge:IA Environment & $0.15$       & $-0.28$  & $0.67$       \\
                                               & $(0.10)$     & $(0.17)$ & $(0.75)$     \\
Social Cue Nudge:IH Environment & $0.16$       & $-0.10$  & $0.60$       \\
                                               & $(0.11)$     & $(0.18)$ & $(0.77)$     \\
\hline
R$^2$                                          & $0.14$       & $0.04$   & $0.01$       \\
Adj. R$^2$                                     & $0.13$       & $0.03$   & $-0.00$      \\
Num. obs.                                      & $355$        & $355$    & $355$        \\
\hline
\multicolumn{4}{l}{\scriptsize{$^{***}p<0.001$; $^{**}p<0.01$; $^{*}p<0.05$}}
\end{tabular}
\end{center}
\caption{Full regression results for the ``Interaction'' Models presented for the three outcomes of interest.}
\label{tab:full-interaction-model}
\end{table*} 

\subsection{E.2: Treatment-on-Treated Models}
In addition to the intent-to-treat analysis, we also conducted a treatment-on-treated analysis for the ``Base'' and ``Interaction'' models. For the ``Social Cue Nudge'' users were only included in the treatment if they actually triggered the intervention, rather than if they were just sorted into the group. Results from the ``Base'' model of this analysis can be found in Figure \label{fig:treatment-on-treated-main-effects}. Results from both the ``Base'' model and the ``Interaction'' model echo our findings from the more conservative intent-to-treat analysis. 

\begin{figure}[t]
    \centering
    \includegraphics[width=1\linewidth]{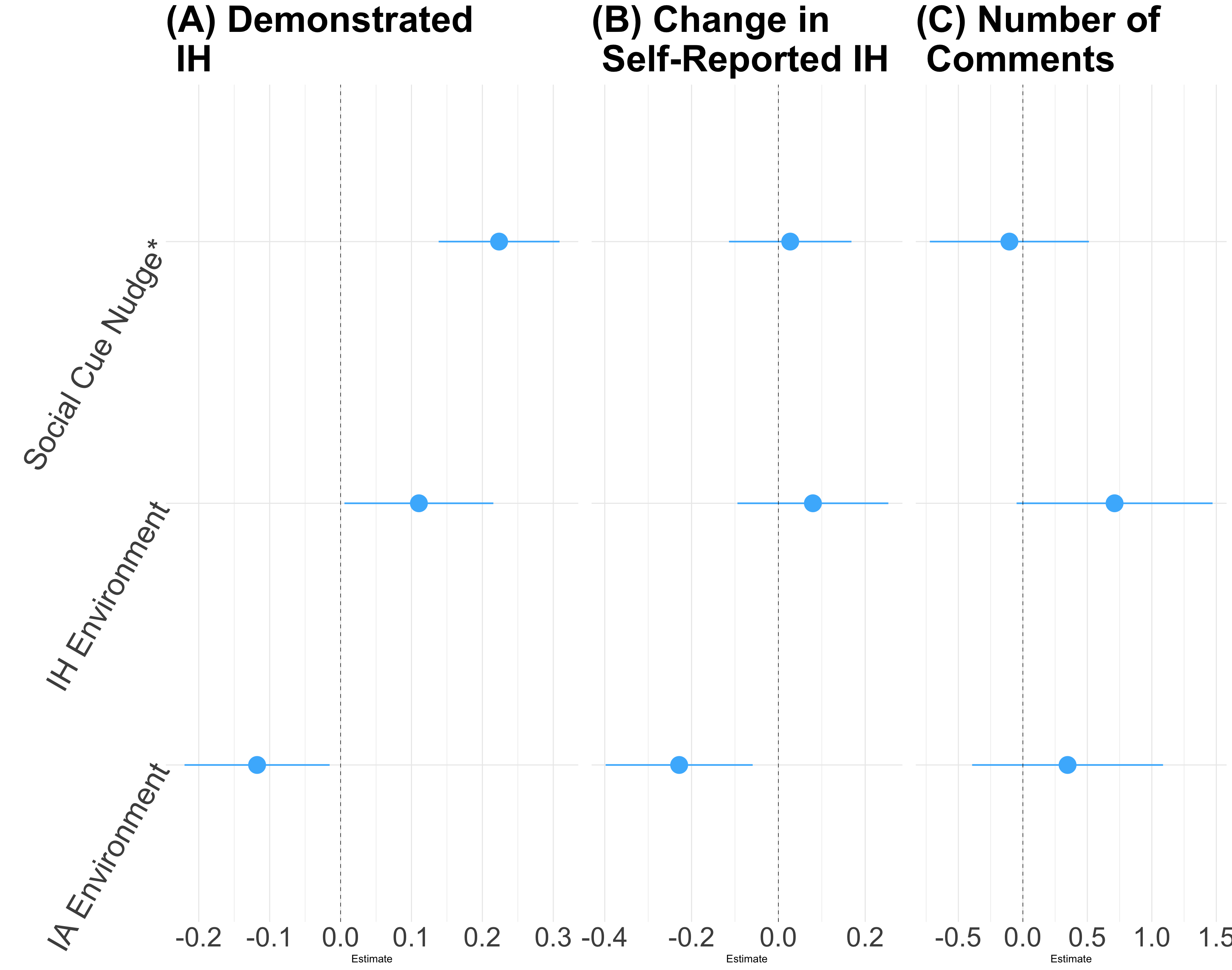}
    \caption{The coefficients for the ``Base'' model shown in Table \ref{tab:modelformulas} with three outcomes of interest. Error bars represent 95\% confidence intervals. Social Cue Nudge* represents only users who actually triggered the Social Cue intervention, as opposed to those who were randomized into the treatment group}
    \label{fig:treatment-on-treated-main-effects}
\end{figure}

\begin{figure}[!]
    \centering
    \includegraphics[width=1\linewidth]{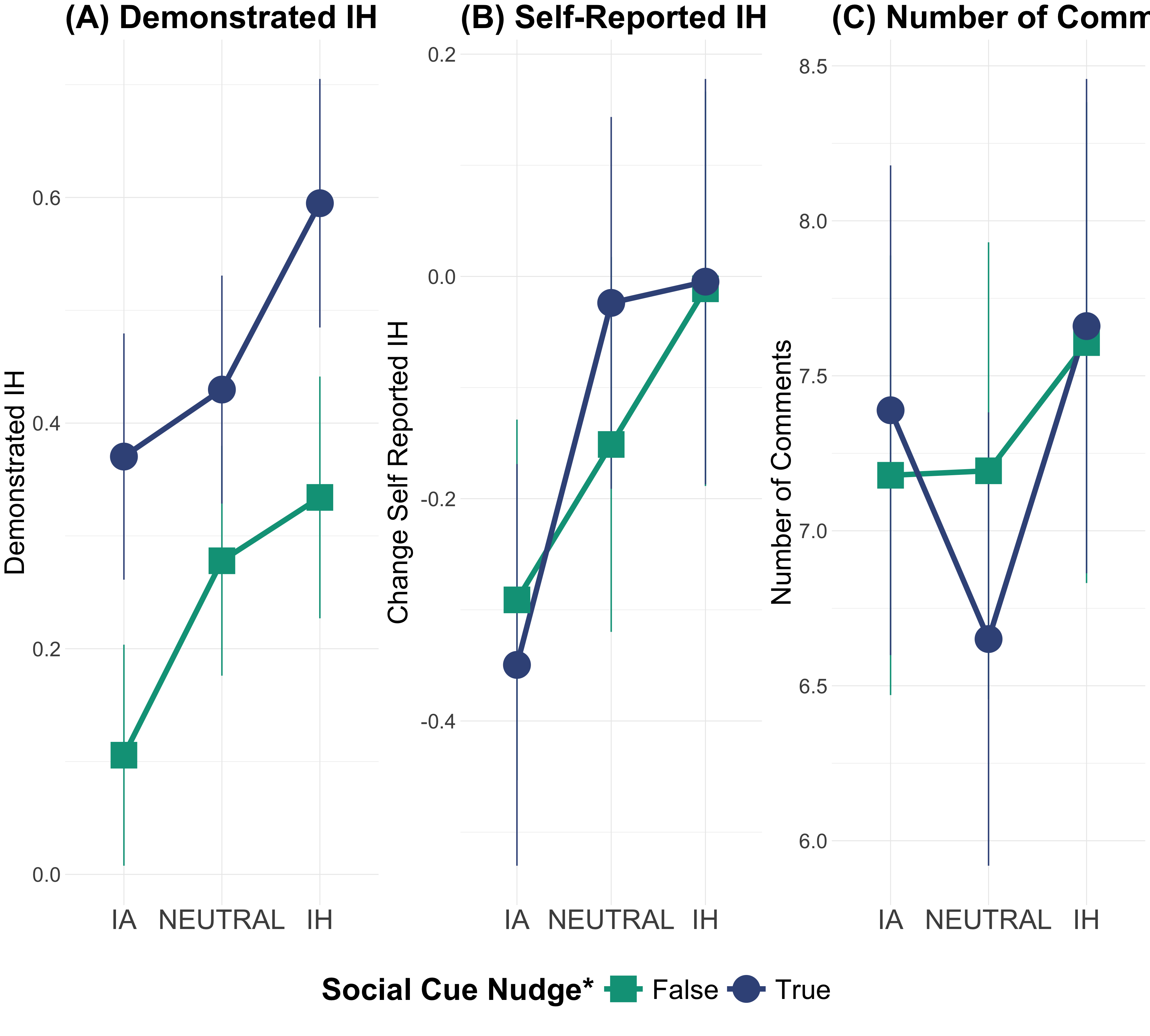}
    \caption{Interactoins between the effects of the treatments for the ``Interaction'' model shown in \ref{tab:modelformulas} Plotted val-
ues are predicted outcomes of interest from each model and
slopes indicate interaction effects. Error bars represent 95\%
confidence intervals.}
    \label{fig:treatment-on-treated-main-effects}
\end{figure}

\subsection{E.3: Covariate Models}
Table \ref{tab:modelformulas-covariates} shows specifications for our pre-registered models ``Base'' and ``Interaction'' models with participant covariates. The full results from the ``Base with Participant Covariates'' is shown in Table \ref{tab:base-w-covars-full}. The full results from the ``Interaction with Participant Covariates'' is shown in Table \ref{table:covariate-coefficients}.
\begin{table*}[ht]
    \centering
    \begin{tabular}{p{2.5in}|p{7in}}
        Model&Formulation\\
        Base with Participant Covariates & $Y_i = \beta_0 + \beta_1 \cdot Cue + \beta_2 \cdot Env + \beta_pP_i$ \\
        Interaction with Participant Covariates & $Y_i = \beta_0 + \beta_1 \cdot Cue + \beta_2 \cdot Env + \beta_I (Cue \cdot Env)$ \\
        & $+ \beta_pP_i + \beta_{P_i, C}(Cue \cdot P_i) + \beta_{P_i, E}(Env \cdot P_i)+ \beta_{P_i, CE}(Cue \cdot Env \cdot P_i)$ \\
    \end{tabular}
    \caption{The formulation for the two covariate models used to evaluate our intervention. $Y_i$ represents the three outcomes of interest - demonstrated IH, self-reported IH, or number of comments; $Cue$ is a binary variable indicating whether or not a participant was assigned to the ``Social Cues” treatment group; Env represents which environment a participant was assigned to (IA, Neutral, or IA); $Cue\cdot Env$ ($CE$) represents the ``Social Cue'' x ``Environment classification'' of a participant $p$; $P_i$ is a matrix containing the relevant participant covariates: the topic-stance group of participant $p$ and $P_{p, 2}$ is the baseline IH of participant $p$. Note that when the outcome is self-reported IH, $P$ does not contain baseline IH as it is used to define the dependent variable in that case.}
    \label{tab:modelformulas-covariates}
\end{table*}

\begin{table*}
\begin{center}
\begin{tabular}{l c c c}
\hline
 & Demonstrated IH & Change in Self Reported IH & Number of Comments \\
\hline
(Intercept)           & $0.30^{***}$  & $-0.17$     & $6.72^{***}$ \\
                      & $(0.05)$      & $(0.09)$    & $(0.42)$     \\
Social Cue Nudge                                               & $0.25^{***}$  & $0.01$      & $-0.10$      \\
                      & $(0.04)$      & $(0.07)$    & $(0.32)$     \\
IA Environment                                              & $-0.12^{*}$   & $-0.22^{*}$ & $0.33$       \\
                      & $(0.05)$      & $(0.09)$    & $(0.38)$     \\
IH Environment                                              & $0.10^{*}$    & $0.09$      & $0.68$       \\
                      & $(0.05)$      & $(0.09)$    & $(0.39)$     \\
Pro-Life                         & $-0.15^{*}$   & $0.13$      & $-0.12$      \\
                      & $(0.06)$      & $(0.11)$    & $(0.49)$     \\
Climate Change (Anti Gov Intervention)                            & $-0.11^{*}$   & $0.20^{*}$  & $0.17$       \\
                      & $(0.06)$      & $(0.10)$    & $(0.43)$     \\
Climate Change (Pro Gov Intervention)                               & $-0.35^{***}$ & $0.15$      & $0.28$       \\
                      & $(0.10)$      & $(0.17)$    & $(0.74)$     \\
Immigration Control (Stricter) & $-0.16^{*}$   & $0.04$      & $0.22$       \\
                      & $(0.08)$      & $(0.14)$    & $(0.60)$     \\
Immigration Control (Looser)    & $-0.28^{***}$ & $-0.02$     & $0.44$       \\
                      & $(0.06)$      & $(0.11)$    & $(0.49)$     \\
Baseline IH       & $0.21^{***}$  &             & $0.49$       \\
                      & $(0.06)$      &             & $(0.45)$     \\
\hline
R$^2$                 & $0.23$        & $0.05$      & $0.02$       \\
Adj. R$^2$            & $0.21$        & $0.03$      & $-0.01$      \\
Num. obs.             & $355$         & $355$       & $355$        \\
\hline
\multicolumn{4}{l}{\scriptsize{$^{***}p<0.001$; $^{**}p<0.01$; $^{*}p<0.05$}}
\end{tabular}
\end{center}
  \caption{Full regression results for the ``Base with Participant Covariates'' Models presented for the three outcomes of interest.} 
  \label{tab:base-w-covars-full} 
\end{table*} 

\begin{table*}
\begin{center}
\begin{tabular}{p{3in} p{1.3in} p{1.3in} p{1.3in}}
\hline
 & Demonstrated IH & Change in Self Reported IH & Number of Comments \\
\hline
(Intercept)           & $0.27^{**}$  & $-0.34^{*}$  & $6.22^{***}$ \\
& $(0.10)$     & $(0.17)$     & $(0.78)$     \\
Social Cue Nudge & $0.25$       & $0.34$       & $0.50$       \\
& $(0.14)$     & $(0.24)$     & $(1.09)$     \\
IA Environment & $0.11$       & $0.38$       & $0.20$       \\
 & $(0.16)$     & $(0.27)$     & $(1.22)$     \\
IH Environment                                              & $0.07$       & $0.45$       & $1.27$       \\
 & $(0.14)$     & $(0.24)$     & $(1.09)$     \\
Pro-Life                         & $-0.28$      & $-0.15$      & $-0.13$      \\
 & $(0.15)$     & $(0.26)$     & $(1.17)$     \\
Climate Change (Anti Gov Intervention)                            & $-0.13$      & $0.07$       & $1.31$       \\
 & $(0.13)$     & $(0.22)$     & $(0.98)$     \\
Climate Change (Pro Gov Intervention)                               & $-0.17$      & $0.31$       & $-0.08$      \\
 & $(0.19)$     & $(0.33)$     & $(1.50)$     \\
Immigration Control (Stricter) & $-0.19$      & $-0.31$      & $-0.35$      \\
 & $(0.24)$     & $(0.41)$     & $(1.84)$     \\
Immigration Control (Looser)    & $-0.16$      & $0.03$       & $0.76$       \\
 & $(0.16)$     & $(0.28)$     & $(1.26)$     \\
Baseline IH       & $0.44^{**}$  & $0.59^{*}$   & $1.73$       \\
 & $(0.16)$     & $(0.28)$     & $(1.26)$     \\
Social Cue Nudge:IA Environment                       & $-0.09$      & $-1.09^{**}$ & $0.17$       \\
 & $(0.20)$     & $(0.35)$     & $(1.57)$     \\
Social Cue Nudge:IH Environment                       & $-0.15$      & $-0.44$      & $0.15$       \\
 & $(0.21)$     & $(0.36)$     & $(1.62)$     \\
Social Cue Nudge:Pro-Life  & $0.11$       & $0.34$       & $-0.28$      \\
 & $(0.21)$     & $(0.37)$     & $(1.65)$     \\
Social Cue Nudge:Climate Change (Anti Gov Intervention)     & $0.06$       & $0.44$       & $-1.57$      \\
 & $(0.18)$     & $(0.31)$     & $(1.40)$     \\
Social Cue Nudge:Climate Change (Pro Gov Intervention)        & $-0.62$      & $-2.40^{**}$ & $-0.70$      \\
 & $(0.45)$     & $(0.77)$     & $(3.46)$     \\
Social Cue Nudge:Immigration Control (Stricter)                         & $-0.01$      & $0.40$       & $0.37$       \\
 & $(0.29)$     & $(0.50)$     & $(2.25)$     \\
Social Cue Nudge:Immigration Control (Looser)                            & $-0.14$      & $-0.10$      & $-0.16$      \\
 & $(0.22)$     & $(0.38)$     & $(1.71)$     \\
Social Cue Nudge:Baseline IH                               & $-0.24$      & $-1.25^{**}$ & $-1.87$      \\
 & $(0.22)$     & $(0.39)$     & $(1.74)$     \\
IA Environment:Pro-Life & $0.10$       & $0.07$       & $2.00$       \\
 & $(0.22)$     & $(0.38)$     & $(1.72)$     \\
IH Environment:Pro-Life & $0.16$       & $0.11$       & $-0.94$      \\
 & $(0.21)$     & $(0.37)$     & $(1.65)$     \\
IA Environment:Climate Change (Anti Gov Intervention)    & $0.01$       & $-0.27$      & $-0.54$      \\
 & $(0.19)$     & $(0.32)$     & $(1.45)$     \\
IH Environment:Climate Change (Anti Gov Intervention)    & $-0.04$      & $-0.04$      & $-1.14$      \\
 & $(0.21)$     & $(0.35)$     & $(1.59)$     \\
IA Environment:Climate Change (Pro Gov Intervention)       & $-0.35$      & $-0.05$      & $1.66$       \\
 & $(0.28)$     & $(0.48)$     & $(2.14)$     \\
IH Environment:Climate Change (Pro Gov Intervention)       & $-0.07$      & $-0.43$      & $1.36$       \\
 & $(0.31)$     & $(0.53)$     & $(2.37)$     \\
IA Environment:Immigration Control (Stricter)                        & $-0.18$      & $-0.03$      & $0.85$       \\
 & $(0.29)$     & $(0.51)$     & $(2.27)$     \\
IH Environment:Immigration Control (Stricter)                        & $0.25$       & $0.33$       & $5.37^{*}$   \\
 & $(0.34)$     & $(0.58)$     & $(2.59)$     \\
IA Environment:Immigration Control (Looser)                           & $-0.23$      & $-0.41$      & $-0.83$      \\
 & $(0.22)$     & $(0.38)$     & $(1.71)$     \\
IH Environment:Immigration Control (Looser)                           & $-0.11$      & $-0.10$      & $-1.78$      \\
 & $(0.22)$     & $(0.38)$     & $(1.70)$     \\
 \end{tabular}
 \end{center}
\end{table*}

\begin{table*}
\begin{center}
\begin{tabular}{p{3in} p{1.3in} p{1.3in} p{1.3in}}
\hline
 & Demonstrated IH & Change in Self Reported IH & Number of Comments \\
\hline
IA Environment:Baseline IH                              & $-0.72^{**}$ & $-1.02^{**}$ & $-1.19$      \\
 & $(0.23)$     & $(0.39)$     & $(1.74)$     \\
IH Environment:Baseline IH                              & $-0.19$      & $-0.97^{**}$ & $-1.29$      \\
 & $(0.21)$     & $(0.36)$     & $(1.63)$     \\
Social Cue Nudge:IA Environment:Pro-Life                         & $-0.16$      & $-0.17$      & $-1.37$      \\
 & $(0.31)$     & $(0.53)$     & $(2.40)$     \\
Social Cue Nudge:IH Environment:Pro-Life                         & $0.14$       & $0.02$       & $-0.14$      \\
 & $(0.32)$     & $(0.54)$     & $(2.43)$     \\
Social Cue Nudge:IA Environment:Climate Change (Anti Gov Intervention)                            & $-0.27$      & $0.39$       & $1.73$       \\
 & $(0.27)$     & $(0.47)$     & $(2.10)$     \\
Social Cue Nudge:IH Environment:Climate Change (Anti Gov Intervention)                            & $0.20$       & $-0.47$      & $-0.12$      \\
 & $(0.28)$     & $(0.49)$     & $(2.18)$     \\
Social Cue Nudge:IA Environment:Climate Change (Pro Gov Intervention)                               & $0.50$       & $2.60^{**}$  & $-0.99$      \\
 & $(0.53)$     & $(0.92)$     & $(4.10)$     \\
Social Cue Nudge:IH Environment:Climate Change (Pro Gov Intervention)                               & $1.39^{*}$   & $2.90^{**}$  & $-2.71$      \\
 & $(0.65)$     & $(1.11)$     & $(4.98)$     \\
Social Cue Nudge:IA Environment:Immigration Control (Stricter) & $0.06$       & $0.13$       & $-1.72$      \\
 & $(0.41)$     & $(0.71)$     & $(3.18)$     \\
Social Cue Nudge:IH Environment:Immigration Control (Stricter) & $0.03$       & $-0.27$      & $-6.46^{*}$  \\
 & $(0.42)$     & $(0.72)$     & $(3.23)$     \\
Social Cue Nudge:IA Environment:Immigration Control (Looser)    & $-0.03$      & $0.76$       & $-0.05$      \\
 & $(0.31)$     & $(0.54)$     & $(2.42)$     \\
Social Cue Nudge:IH Environment:Immigration Control (Looser)    & $0.39$       & $-0.04$      & $3.67$       \\
 & $(0.31)$     & $(0.54)$     & $(2.42)$     \\
Social Cue Nudge:IA Environment:Baseline IH       & $0.78^{**}$  & $2.18^{***}$ & $1.93$       \\
 & $(0.30)$     & $(0.52)$     & $(2.32)$     \\
Social Cue Nudge:IH Environment:Baseline IH       & $0.33$       & $1.52^{**}$  & $1.27$       \\
 & $(0.30)$     & $(0.52)$     & $(2.33)$     \\
\hline
R$^2$                 & $0.31$       & $0.18$       & $0.12$       \\
Adj. R$^2$            & $0.22$       & $0.08$       & $0.00$       \\
Num. obs.             & $355$        & $355$        & $355$        \\
\hline
\multicolumn{4}{l}{\scriptsize{$^{***}p<0.001$; $^{**}p<0.01$; $^{*}p<0.05$}}
\end{tabular}
\caption{Full regression results from ``Full with Participant Covariates" Models presented for the three outcomes of interest.}
\label{table:covariate-coefficients}
\end{center}
\end{table*}

\end{document}